\documentclass[preprint2]{aastex}  

\newcommand{\II}{\protect\small II \normalsize $\!\!$}
\newcommand{\III}{\protect\small III \normalsize $\!\!$}
\newcommand{\HII}{\mbox{\rm H\II}}
\newcommand{\Htwo}{${\rm H_2}$}
\newcommand{\I}{\protect\small I \normalsize $\!\!$}
\newcommand{\HI}{\mbox{\rm H\I}}

\newcommand{\cii}{\mbox{\rm [C\II}]}

\newcommand{\UITunits}{$\rm{ergs~cm^{-2} s^{-1} \AA^{-1}}$}

\slugcomment{To appear in the Astrophysical Journal, Vol. 538}

\shorttitle{A New Probe of the Molecular Gas in M101}
\shortauthors{Smith, Allen, Bohlin, Nicholson, \& Stecher}

\begin{document}

\title{A New Probe of the Molecular Gas in Galaxies:\\
Application to M101}

\author{Denise A.~Smith, Ronald J.~Allen, and Ralph C.~Bohlin}
\affil{Space Telescope Science Institute, 3700 San Martin Drive, Baltimore, 
MD 21218}
\email{dsmith@stsci.edu,rjallen@stsci.edu}
\author{Natalya Nicholson}
\affil{McGill University, Montreal, Quebec, Canada}

\and

\author{Theodore P.~Stecher}
\affil{NASA Goddard Space Flight Center, Laboratory for Astronomy \& Solar 
Physics,\\
Code 680, Greenbelt, MD 20771}


\begin{abstract}

Recent studies of nearby spiral galaxies suggest that photodissociation
regions (PDRs) are capable of producing much of the observed \HI\ in
galaxy disks.  In that case, measurements of the observed \HI\ column
density and the far--ultraviolet (FUV) photon flux responsible for the
photodissociation process provide a new probe of the volume density of
the local underlying molecular hydrogen.  We develop the method and
apply it to the giant Scd spiral M101 (NGC~5457).  The \HI\ column density and
amount of FUV emission have been measured for a sample of 35
candidate PDRs located throughout the disk of M101 using the Very
Large Array and the Ultraviolet Imaging Telescope. We find that, after
correction for the best-estimate gradient of metallicity in the ISM of M101 and
for the extinction of the ultraviolet emission, molecular gas with a
narrow range of density from 30--1000 cm$^{-3}$ is found near
star-forming regions at all radii in the disk of M101 out to a distance
of $12' \approx 26$ kpc, close to the photometric limit of $R_{25}
\approx 13.5'$.

In this picture, the ISM is virtually all molecular in the inner parts
of M101.  The strong decrease of the \HI\ column density in the inner
disk of the galaxy at $R_G < 10$ kpc is a consequence of a
strong increase in the dust-to-gas ratio there, resulting in an
increase of the \Htwo\ formation rate on grains and a corresponding
disappearance of hydrogen in its atomic form.

\end{abstract}


\section{Introduction}
\label{sec:intro}

Massive gas clouds in the ISM are considered to be the progenitors of
young stars in the conventional picture of star formation in galaxy
disks.  The gas may initially be either in an atomic or in a molecular
state \citep{elm95} depending on the ambient physical conditions (UV
flux and gas pressure), but as the volume densities increase through
self-gravity, the gas becomes mostly molecular on its way to forming
stars. Once the gas is molecular and at densities above $\approx
100\:{\rm cm}^{-3}$, the cooling rate is so high (e.g.~\citet{gol78})
that rapid collapse follows.  The rate-determining step in the
conventional picture is then the aggregation of \HI\ or diffuse \Htwo\
and the formation of Giant Molecular Clouds. \citet{elm93} has
discussed this problem in detail.

For many late-type systems, the rate-determining step along the path
to star formation appears to be the conversion of \HI\ clouds into
\Htwo.  The fraction of the ISM in atomic form is thought to increase
as one progresses along the Hubble Sequence towards the late-type
galaxies (see e.g.\ the review by \citet{rob94}).  Furthermore, the
atomic fraction is thought to dominate the molecular by an order of
magnitude or more for a significant number of actively-star-forming
galaxies of types Sc and later \citep{you89}.

However, in recent years evidence has been accumulating that a
significant fraction of the \HI\ in the ISM of galaxies is in fact not
a precursor to the star formation process, but instead is produced by
photodissociation of the \Htwo\ by UV photons emanating from nearby
newly-formed massive stars. In that case, either the way we compute the
molecular fraction in galaxies via the CO(1-0) emission is wrong, or
our ideas of how to control the otherwise runaway collapse of molecular
clouds to form stars need further modification.  \citet{shu97} has
summarized the view that magnetic fields may be a controlling factor in
cloud collapse.

The dissociation process was first described by \citet{ste67}.  
The \Htwo\ absorbs photons emitted at 1108 \AA\ and 1008 \AA\
via electronic transitions in the Lyman ($X ^1\Sigma_g^+ \rightarrow B
^1\Sigma_u^+$) and Werner ($X ^1\Sigma_g^+ \rightarrow C ^1\Pi_u^+ $)
bands.  In the subsequent decay to the vibrationally excited levels of
the ground electronic state, $\sim 10$\% of the \Htwo\ molecules will
dissociate.  Photons with wavelengths as long as $\approx 1850$ \AA\
can continue to create \HI\ by dissociating the ``pumped'' ($X
^1\Sigma_g^+, 0 < v < 14$) \Htwo\ via additional Lyman and Werner band
transitions.  The UV fluorescence spectrum predicted from this process was
first observed by \citet{wit89} in the nebula IC~63.  

The initial evidence that \HI\ may be a \textit{product} of the star
formation process rather than a precursor to it was discovered in a
comparison of the morphology of dust lanes and \HI\ ridges in a
prominent spiral arm of the galaxy M83 (NGC 5236) by
\citet{all85,all86}. This comparison showed that the ridges defining
the \HI\ spiral arms peaked not on the dust lanes, as would have been
expected in the conventional picture of density-wave-triggered star
formation \citep{rob69}, but instead they were co-linear with the
chain of \HII\ regions created by young massive stars forming
\textit{downstream} from the dust lane.  Other studies followed on M83
and other nearby spiral galaxies and have generally reached the same
conclusions: for M83 \citep{til93} , for M51
\citep{vog88,til87,til88,til89,til90,til91,ran92,kna92}, and for M100
\citep{kna94,kna96}.

Infrared data are providing important support for this picture. For
instance, models constrained by KAO measurements of the $158\:\mu$m
\cii\ line indicate that photodissociation may produce as much as
70\%--80\% of the \HI\ in the disk of NGC~6946 \citep{mad93}.  More
recently, spectra of nearby galaxies taken with the ISO satellite
indicate that the bulk of the mid-infrared emission from galaxy disks
arises in photodissociation regions (PDRs) \citep{lau99,rou99,vig99}.

Photodissociation therefore appears to be an important, and in some
cases perhaps even the dominant, factor in defining the appearance and
distribution of \HI\ in galaxies, at least over the main body of their
stellar disks.  In the PDR picture, \HI\ is produced on the surfaces
of molecular clouds by the action of far-UV photons from nearby O and
B (and some A) stars.  The physics of PDRs has been worked out in
detail (see e.g.\ the review by \citet{hol99}), partly in response to
new observational results on the infrared spectra of nearby Galactic
star-forming regions such as Orion.

Recognition of the layered structure of the different atomic and
molecular species which exist in dense, bright PDRs like the Orion Bar
($n \approx 5 \times 10^4\:{\rm cm}^{-3}$, \citet{tie93})
requires linear resolution better than $\sim 0.02$ pc, and the
\HI\ layer in this case would be only $\approx 0.015$ pc thick. However,
PDRs are also formed by modest UV fluxes incident on low-density clouds,
and the \HI\ layers in this case are much thicker. 
For example, \citet{wil96} have identified a
large, low-density PDR in the Galaxy, near the molecular cloud G216-2.5. 
In this case, the \HI\ layer is probably 10-50 pc thick ($n \approx
50-10\;{\rm cm}^{-3}$), and appears in projection as an object about
$50 \times 200$ pc in size. While the IR emission from such large,
low-density PDRs is hard to detect (but see \citet{luh96}), the
\HI\ emission is actually easier to measure, since the PDR is 
larger and therefore covers a larger fraction of the radio telescope 
beam at $\lambda$21 cm. The problem with such measurements in the
Galaxy which Williams and Maddelena had to circumvent is the confusion of
the \HI\ profiles by unrelated gas along the same line of sight.
Synthesis imaging radio telescopes can now offer linear resolutions
of order 100--200 pc in the nearby galaxies, and the confusion problem can be
minimized by choosing galaxies which are viewed more face-on.

The combination of high spatial resolution ($\sim 100-200$ pc) \HI\
mapping along with far ultraviolet (FUV) and H$\alpha$ imagery
provides a unique opportunity to explore the details of the
photodissociation picture in external galaxies.  FUV imagery at
$\approx 1500$ \AA\ ($\approx 8.3$ eV) directly measures the lower
energy photons which can dissociate ``pumped'' molecular gas, and
serves as a tracer for the more energetic photons in the range $912 <
\lambda < 1108$ \AA\ which start the pumping process for ground-state
\Htwo.  H$\alpha$ imagery is used to assess the importance of
extinction, which may be substantial in the far
ultraviolet. \citet{all97}, hereafter A97, have used this technique
with the Sb(r)I-II galaxy M81, finding that patches of \HI\ emission
are often located at the periphery of star forming regions mapped in
the FUV, as expected for PDRs.  The \HI\ column densities in M81 are
consistent with the photodissociation of a low density molecular gas
by relatively modest UV fluxes\footnote{However, A97 did not adjust
their results for the possible effects of internal extinction of the
UV flux in M81, nor for changes in the dust-to-gas ratio. As we shall
see in this paper, both corrections can be important.}.

In this paper, we extend the work of A97 to the Sc(s)I galaxy M101 and
carry out a much more extensive and quantitative analysis of the data.
Owing to its proximity (7.4 Mpc, \citet{kel96}), its nearly face--on
orientation, and the availability of FUV, H$\alpha$, and \HI\ data,
M101 is the next obvious candidate for this type of work.  The best
angular resolution presently available in \HI\ for M101 is $\approx
6''$; this corresponds to $\approx 215$\ pc, just adequate to permit
identification of the larger PDRs, but still not sufficient to provide
much morphological detail.  The M101 data are discussed in \S
\ref{sec:data}; they have all been drawn from existing archival data
sets.  A representative sample of PDRs is identified and discussed in \S
\ref{sec:results}.  The volume density of the gas, $n$, associated
with each PDR is derived in \S \ref{sec:volume}. The implications of
the resulting radial profile of $n$ are discussed in \S \ref{sec:imp}.

The method used in this paper provides a new probe for studying the
molecular gas content of galaxies which can be compared with the
conventional results obtained from using the CO molecular emission
(e.g.~\citet{sol83}).  Questions have been raised about the
reliability of CO as a tracer for molecular hydrogen in galaxy disks
(e.g.~\citet{all96}; \citet{com99}), and it is of interest to explore
other ways of establishing the amount and physical state of the ISM in
different parts of galaxies.

\section{Archival Data Used}
\label{sec:data}

The FUV image of M101 was obtained by the Ultraviolet Imaging
Telescope (UIT) \citep{ste92,ste97} during the Astro-2 mission with
the B1 filter ($\lambda=1520$ \AA, $\Delta \lambda=354$ \AA) and an
exposure time of 1310 s.  The spatial resolution and pixel spacing of
the image are $3''$ FWHM and $1.13''$, respectively.  Astrometric and
photometric uncertainties are $\sigma = 1.8''$ and $\sim 15$\%,
respectively.  An image of the H$\alpha$ emission with resolution
$\approx 2''$ FWHM was obtained from L.\ van Zee and M.\ Haynes.  The
VLA 21 cm line data were provided by R.\ Braun. The spatial resolution
is $6''$ FWHM (215 pc), and the pixel spacing is $2''$ (72 pc).  The
velocity channels are separated by 5.16 km s$^{-1}$ and the velocity
resolution is 6.2 km s$^{-1}$.  We re-derived a map of the \HI\ column
density, $N$(\HI), by using the pixels in the channel maps whose
values exceed 3 times the $1 \sigma$ noise value of 16.8 K.  The
radial profile of the \HI\ column density derived from our data agrees
with the original data portrayed in Figure 8a of \citet{bra97}.
Full-size images of the FUV, H$\alpha$, and \HI\ emission from M101
can be found in \citet{wal97}, \citet{zee98}, and \citet{bra95},
respectively.

The original astrometry of the FUV and H$\alpha$ images appears to be
excellent based upon the general agreement in the positions of FUV
peaks and H$\alpha$ sources, and in the large--scale spatial
distribution of the FUV continuum and the H$\alpha$ and 21~cm line
emission. The astrometry of the \HI\ images is intrinsically $\lesssim
1''$, owing to the calibration procedures used at the VLA. The FUV and
H$\alpha$ images are then smoothed and resampled to match the spatial
resolution and pixel spacing of the \HI\ data.

\section{Results and Analysis}
\label{sec:results}

The relationship between the FUV, H$\alpha$, and \HI\ emission is
illustrated in Figures \ref{fig:overlay1} and \ref{fig:overlay2}.
Boxes indicate the locations of regions shown in more detail in Figures
\ref{fig:fuvha2} through \ref{fig:fuvhi3}.  These regions are chosen
arbitrarily to provide examples of the detailed FUV, H$\alpha$,
and \HI\ morphology.  A large--scale spatial correspondence between the
21 cm line emission and the FUV and H$\alpha$ emission is observed, as
expected in both the conventional and photodissociation pictures.  Star
formation occurs where ``sufficient amounts'' of gas have accumulated
in the conventional picture. However, that picture does not yet provide a
\textit{quantitative} connection between e.g. the amount of precursor
gas (both atomic and molecular) which is present and the number of
young stars formed from it. Determining this relation from the
observations is currently an active area of research in the study of
star formation in galaxies (e.g.\ \citet{ken98}).

On the other hand, the photodissociation picture provides a direct,
quantitative connection between the FUV flux emanating from a
region of recent star formation, the volume density of the \Htwo, and
the column density of the surrounding \HI. We have chosen in what
follows to interpret our results in terms of the photodissociation
picture. While we clearly favor such an approach, we must emphasize
that our data, by themselves, do not provide a definitive argument
against the conventional picture.

\subsection{FUV and H$\alpha$: Mapping Star Forming Regions}
\label{sec:pdrs_fuvha}

Extinction effects are much stronger at 1500 \AA\ than in the visual,
as indicated by reddening curves for the Milky Way (e.g.~$A_{1500}=2.6
A_V$ \citep{sav79}) and the Small Magellanic Cloud (e.g.~$A_{1500}=5.6
A_V$ \citep{hut82}).  This fact raises the concern that
ultraviolet--based studies may be biased if entire star forming
regions are totally obscured at FUV wavelengths.  For this reason, A97
explore the effects of extinction in M81 by comparing the FUV and
H$\alpha$ emission.  These authors find a strong correspondence
between the small--scale structure of the FUV and H$\alpha$ emission
in M81, with every bright, reliable H$\alpha$ source having a FUV
counterpart.  This suggests that the FUV morphology is relatively
unaffected by extinction, or at least that the FUV and the H$\alpha$
are affected in the same way.

To explore the importance of extinction in M101, we examined a set of
non--stellar H$\alpha$ sources for FUV counterparts.  We selected
objects with the simple goal of constructing a set of representative
sources distributed across the galaxy which could be examined
interactively for FUV counterparts.  The automated point source
extraction routines found in GIPSY and IRAF are sufficient for this
purpose, yielding $\approx 150$ H$\alpha$ sources when using a peak
intensity threshold of 15$\sigma$.  The resulting set does not form a
statistically complete sample.  Visual inspection of the data reveals
that approximately 10\% of the sources lack a detectable FUV
counterpart within a radius of $10''$ (360 pc), at the present
sensitivity.  A comparison with the H$\alpha$ images of M101 presented
in \citet{hod90} shows that these sources are in fact real \HII\
regions, and not artifacts.  Examples of H$\alpha$ sources without FUV
counterparts may be seen in Figure \ref{fig:fuvha2}; these sources
tend to be fainter H$\alpha$ peaks.  Source coordinates are given in
Table \ref{tab:nofuv}.  Sources such as those observed at RA = $14^h
2^m 9.96^s$, Dec$=+54^\circ 35' 16.5''$, and at RA = $14^h 2^m
0.71^s$, Dec = $+54^\circ 32' 2.7''$ are foreground stars, based upon
the continuum and narrow--band imagery of \citet{hod90}.  The
existence of equally faint H$\alpha$ sources with FUV counterparts
indicates that the lack of FUV emission is more likely to be an
extinction effect than a sensitivity issue.

We conclude that FUV wavelengths effectively map the locations of the
majority of the star forming regions in M101, without {\it severe}
extinction effects.  (The amount of extinction will be discussed
further in Section \ref{sec:tau}). In this case, FUV mapping offers an
additional advantage over H$\alpha$ mapping in that FUV wavelengths
will trace older sites of star formation in which ionizing stars are
no longer present.  Such FUV--bright/H$\alpha$--dim sites are powered
primarily by B stars, whose photon energies are still favorable for
the dissociation of \Htwo.  A97 find that 10\% of 144 FUV sources in
M81 do not exhibit H$\alpha$ emission.  We identified $\approx 350$ FUV
sources distributed across M101, using a threshold peak intensity of
$3 \sigma$.  Of these, the brightest 177 (half of the sources) were
visually examined for the presence of H$\alpha$ counterparts.
Approximately 7\% of the 177 sources do not appear to have an
H$\alpha$ counterpart within $10''$ (360 pc), at the present
sensitivity.  Examples of such sources are seen in Figure
\ref{fig:fuvha3} at A) RA(1950) = $14^h0^m40.28^s$, Dec(1950) =
$+54^\circ 31' 42.1''$; B) RA = $14^h 1^m 0.62^s$, Dec = $+54^\circ
31' 53.7''$; C) RA = $14^h 1^m 3.45^s$, Dec = $+54^\circ 32' 45.3''$.

In the remainder of the paper, we focus our attention on a smaller
sample of approximately 100 FUV sources.  The selection criteria were
driven by our desire to identify a set of representative sources
distributed across M101, manageable in size.  The resulting sample
is not intended to be statistically complete.  The sample contains all
FUV sources with peak intensities $\ge 10 \sigma$ identified by the
point source extraction routines.  These sources comprise
2/3 of the set.  The remaining sources were chosen by eye
to bring the total number of sources to $\approx 100$, while keeping
the distribution uniform across the galaxy.  This technique ensures
that we do not exclude sources that may have been missed by the point
source extractors.  The added sources have peak intensities of $3-10
\sigma$.  All but one (a weak elongated source) belong to the set of
350 sources discussed above.  The properties of the H$\alpha$ and \HI\
emission are not considered during the source selection.

\subsection{The Detailed Morphology of the FUV and \HI}
\label{sec:pdrs_fuvhi}

The small--scale relationship between the FUV and the \HI\ emission is
shown in Figures \ref{fig:fuvhi2} and \ref{fig:fuvhi3}.  The general
distribution of the \HI\ gas follows that of the diffuse FUV emission,
with enhancements in the 21 cm line emission seen near FUV peaks, as
in M81.  All but 2 of the $\approx 100$ FUV sources exhibit 21 cm line
emission within 800 pc of the source.  Specific regions of
interest in M101 include the area surrounding SN~1951, located at
RA(1950) = $14^h02^m07.5^s$, Dec(1950) = $+54^\circ 36' 04''$, at the
inner edge of the prominent eastern spiral arm.  Here, FUV emission is
coincident with and extends from the site of the NGC~5462 \HII\ region
(RA(1950) = $14^h02^m06.8^s$, Dec(1950) = $+54^\circ 36' 20''$, our
Figure \ref{fig:fuvhi2}, see also Figure 7 of \citet{via81}).
Emission from the 21 cm line lies along the periphery of the \HII\
region but is absent in the vicinity of SN1951 itself.  Within the
framework of the photodissociation picture, an explanation
for this morphology could be that this is a region of intense past and
present star formation, as evidenced by the FUV, H$\alpha$ and radio
continuum emission.  Generations of supernovae prior to SN1951 may have
evacuated much of the surrounding molecular gas, resulting in the
apparent hole in the gas distribution.  Non--ionizing photons may have
dissociated the skins of the remaining molecular gas clouds, creating
the observed \HI\ which neatly curves around the existing FUV--emitting
regions.  A sketch of the spatial relationships between the FUV,
H$\alpha$, and \HI\ emission in this type of situation may be found in
Case IV, Figure 8 of A97.

A case in which the star forming regions may lie above the mid--plane
of the galaxy is seen at RA = $14^h 1^m 55.68^s$, Dec = $+54^\circ 33'
23.7''$ in Figure \ref{fig:fuvhi2}. Here, the NGC~5461 complex shows
enhanced \HI\ emission slightly offset from the main FUV source, as
well as spatially coincident \HI.  The NGC~5447 complex, with peaks at
RA = $14^h 0^m 43.07^s$ Dec$=+54^\circ 30' 35.5''$ and RA = $14^h 0^m
41.69^s$ and Dec = $+54^\circ 30' 47.4''$, in Figure \ref{fig:fuvhi3}
is another example of this morphology, which occurs when \HI\ appears
spread over the FUV-emitting region but without much associated
extinction. The \HI\ in this case is behind the FUV source along the
line of sight, as in Case I, Figure 8 of A97.

Another interesting morphological behavior is exemplified by the small
FUV--bright/H$\alpha$--dim source located in the \HI\ depression at
RA = $14^h 0^m 40.30^s$ Dec = $+54^\circ 31' 41.4''$,
just north of NGC~5447 in Figure \ref{fig:fuvhi3}. This site may
represent an older star forming region which has blown a hole through
the plane of the galaxy.  Non--ionizing photons from the remaining B
stars could be dissociating the walls of the hole, producing the
observed \HI.  This scenario corresponds to that illustrated in Case
III, Figure 8 of A97.

Finally, we note that the large loop of \HI\ centered at approximately
RA = $14^h1^m42^s$, Dec = $+54^\circ 31' 45''$ in
Figure \ref{fig:fuvhi2} presents a case where some of the PDRs may be
obscured.  The northern portion of the loop shows the presence of
fainter H$\alpha$ sources and only weak FUV emission.  The southern
portion of the loop contains multiple, brighter FUV sources, however.
Some of these sources also appear to lack H$\alpha$ counterparts, and
may be older sites of star formation (e.g.~the FUV peak located at
RA = $14^h 1^m 39.12^s$, Dec = $+54^\circ 31' 31.9''$).
This behavior could be explained if star forming regions in the
northern portion of the loop lie closer to the mid--plane of the
galaxy, in a behavior intermediate between Cases I and II in Figure 8
of A97.  The northern sources may also be younger
sites of star formation which have not had sufficient time to break
through the ISM.  These types of sources are clearly in the minority,
however; the locations of the majority of candidate PDRs are
effectively mapped by the FUV emission.

\subsection{Candidate PDRs}
\label{sec:35pdrs}

We have used the UIT data to select a group of candidate PDRs for
further study.  Of the $\approx 100$ FUV--bright sources, we
identified 35 regions for which 1) the total FUV flux and 2) the
distance between the FUV source and the peaks of the associated 21 cm
line emission could be reliably measured.  The selected sources are
therefore relatively isolated and in areas containing detectable \HI.
The primary selection criterion is in fact confusion, as 95\% of the
rejections were based upon the presence of another significant FUV
source within $\lesssim 15''$ (540 pc).  Two sources were rejected due
to a lack of 21 cm emission within a radius of $\approx 15''$; one
additional source was rejected due to the weakness of the \HI\
emission.  The locations of the 35 selected regions are shown in
Figure \ref{fig:fuv}.  The regions are classified into two subsets
according to the intensity $\chi$ of the FUV flux seen by the
neighboring \HI\ clouds. The \HI\ column density is directly related
to $\chi$ in the photodissociation picture (see Section
\ref{sec:physics}).  Regions in Figure \ref{fig:fuv} marked with small
square boxes are characterized by a narrow range of $\chi$ such that
$0.9 < \chi < 10$. These regions represent a set of ``standard UV
fluxes.''  Regions designated by open circles have FUV fluxes outside
of this range.  The FUV, H$\alpha$, and \HI\ morphologies of typical
candidate PDRs are illustrated in Figures \ref{fig:pdrfuvha} through
\ref{fig:pdra}.

The quantity $\chi$ is given by $\chi=F_{FUV}(7.4 \times
10^6/\rho_{HI})^2/F_0$, where $F_0=2.64\times 10^{-6}$ ergs cm$^{-2}$ 
s$^{-1}$ \AA$^{-1}$ is
the solar neighborhood value at 1500 \AA\ (A97, and references
therein), $F_{FUV}$ is the total FUV flux observed by UIT in
\UITunits, $\rho_{HI}$ is the distance between the FUV source and the
peak of the neighboring \HI\ emission in parsecs, and $7.4 \times
10^6$ is the adopted distance to M101 (in parsecs).  The value of
$F_0$ is based upon the values of the local interstellar radiation
field given by \citet{dra78} and illustrated in Figure 1 of
\citet{dis88}. The solar neighborhood value varies weakly over the
wavelength range of the UIT B1 filter; the adopted value of $F_0$
would change by less than 10\% if we were to convolve the
\citet{dra78} curve with the UIT B1 filter profile.  

The measurements of $\rho_{HI}$ and $N_{HI}$, the associated column
density, are dependent upon the morphology of the source.  As seen in
Figure \ref{fig:pdra}, multiple patches of 21 cm emission neighbor
each source.  Each patch serves as an equally valid probe of the
surrounding ISM.  We determine $\rho_{HI}$ by calculating the average
\HI\ surface brightness in successive annuli for each PDR.  The annuli
are centered upon the FUV source.  The quantity $\rho_{HI}$ is defined
as the radius at which the surface brightness profile first reaches a
local maximum.  To measure the associated column density, $N_{HI}$, we
identify the brightest patch of 21 cm line emission located at a
distance $\rho_{HI}$ from the FUV source.  The adopted value of
$N_{HI}$ corresponds to the brightest pixel within the patch.  The
distance between the brightest pixel and the FUV source is always
within 0.5 pixels of $\rho_{HI}$.  This measurement of $N_{HI}$ is
preferred to the average value of $N_{HI}$ obtained from the surface
brightness profiles since the annuli include regions lacking 21 cm
line emission.  The average values therefore underestimate the
column densities in the local ISM.

The measurements of $F_{FUV}$, $\rho_{HI}$, $\chi$, and $N_{HI}$ for
each region are listed in Table \ref{tab:data} and displayed in
Figures \ref{fig:fuvflux} through \ref{fig:nhi} respectively.  Filled
circles represent the regions with $0.9 < \chi < 10$.  The values of
the sky background and any surrounding diffuse FUV emission have been
removed from the measurements of $F_{FUV}$.  This is accomplished by
measuring the average FUV surface brightness in successive annuli
centered on the FUV source.  The amount of diffuse $+$ sky background
per unit area is measured at the radius at which the FUV surface
brightness levels off in the resulting radial profile.  This value is
subtracted from the total FUV flux contained within this radius.  The
FUV fluxes (Figure \ref{fig:fuvflux}) vary by a factor of 100, but
show no clear radial trend.  With regard to the correspondence between 
the FUV and \HI\ emission,
Figures \ref{fig:pdra} and \ref{fig:rhohi} show many cases in which
the small--scale \HI\ emission is slightly offset from or surrounds a
FUV peak.  Roughly half (16/35) of the FUV sources display an
enhancement in the \HI\ emission within 250 pc.  For another 16
sources, the separation between the FUV and \HI\ peaks ranges from 250
pc to $\sim 500$ pc.  While the photodissociation scenario suggests
that the \HI\ enhancements are associated with the FUV emission, we
remind the reader that other interpretations of the data may be valid,
and that the observed morphology may simply reflect the general
distribution of \HI\ enhancements in M101.  The \HI\ enhancements for
the remaining 3 sources are fairly distant from the FUV sources
($\rho_{HI}> 500$ pc) and may not be associated with the FUV source.
The values of $\rho_{\HI}$ do not display a significant radial trend 
(Figure \ref{fig:rhohi}).
The resulting values of $\chi$ cover a large range (Figure
\ref{fig:chi}), with no obvious trends with radius in the data.
Values of $N_{HI}$ represent the peak column density associated with
each source.  Figure \ref{fig:nhi} indicates that the values of
$N_{HI}$ cover a wide range and tend to increase with increasing
radius.  This situation reflects the central deficit of 21 cm line
emission in M101, a feature which is commonly observed in galaxies.

Finally, the projected separation between the FUV and H$\alpha$
sources ($\rho_{H\alpha}$) is given (Figure \ref{fig:rhoha}) in order 
to further quantify the effects
of extinction discussed in \S \ref{sec:pdrs_fuvha}.  The majority of the FUV
sources (85\%, 30/35) have an H$\alpha$ counterpart located within
$6''$ (215 pc; the FWHM of the \HI\ image), as seen in Figure
\ref{fig:rhoha}.  Of the remaining sources, 4 have H$\alpha$
emission which is slightly offset (215 pc $< \rho_{H\alpha} < $ 315
pc).  The remaining FUV source lacks an obvious nearby H$\alpha$
counterpart, with the nearest H$\alpha$ source located over 600 pc
away.  This source may be an older star forming region.  The
separation between the FUV and H$\alpha$ emission is independent of
the distance from the center of M101.

\subsection{The Quantitative Link Between the FUV and the \HI}
\label{sec:physics}

The column density of \HI\ produced in a PDR is fundamentally linked
to the amount of far--ultraviolet (FUV) emission incident upon it
and the local molecular gas volume density. The production
of \HI\ is calculated with the same physics as the production of the
H$_2$ near--infrared emission lines by fluorescence. Following e.g.\ 
\citet{ste88},
\begin{equation}
N(\HI)={1 \over \sigma} \ln{\left[ {{D G} \over {R n}}\chi  + 1 \right]},
\label{eqn:nhi}
\end{equation}

\noindent where \\

\begin{tabular}{rcp{1.8in}}
$N(\HI)$ & = & the \HI\ column density, \\
$\sigma$ & = & the effective grain absorption cross section per H nucleus, \\
$D$      & = & the average unattenuated \Htwo\ photodissociation rate, \\
$R$      & = & the \Htwo\ formation rate coefficient, \\
$\chi$   & = & the FUV intensity relative to the local ISRF, and \\
$n$      & = & the total volume density of the gas. \\
\end{tabular}

\noindent The quantity $G$ is a dimensionless function of $\sigma$,
the absorption self--shielding 
function $f$, and the column density of molecular hydrogen $N_{{\rm H2}}$: 
\begin{displaymath}
G = \int_0^{N_{{\rm H2}}} \sigma f e^{-2\sigma N_{{\rm H2}}^\prime} {\rm d}N_{{\rm H2}}^\prime \rightarrow \mbox{constant}.
\end{displaymath}
$G$ becomes constant for large values of $N_{{\rm H2}}$ due to
self--shielding \citep{ste88}.

Our analysis ignores the ionized component of the \HI,
which is produced by photons of higher energy ($\geq 13.6$ eV) than
those we measure ($\approx 8.3$ eV). If such photons are present in the
flux impinging on the molecular cloud, then some of the \HI\ would
disappear into \HII\ and we would be underestimating $N(\HI)$. However,
in general both the PDR models and the highest-resolution observations
in the Galaxy (e.g.\ figures 3 and 30 in \citet{hol99}) show that most
of the \HII\ appears closer to the exciting stars than does the \HI, so
any correction ought to be small.

Equation \ref{eqn:nhi} is strongly dependent upon the dust--to--gas ratio
($\delta=A_V/N_H$) and weakly dependent upon the gas temperature ($T$), since
\begin{eqnarray}
\sigma & = & 1.883 \times 10^{-21} (\delta/\delta_0)\ {\rm cm}^{-2}, \nonumber\\
R      & = & 3.0 \times 10^{-18} (\delta/\delta_0) T^{1/2} y_F(T)\ {\rm cm^3 s^{-1}}, \nonumber\\
G      & = & (\sigma/\sigma_0)^{1/2}G_0, \nonumber
\end{eqnarray}
where $\delta_0$ and $\sigma_0$ refer to values in the solar neighborhood,
and $y_F(T)$ represents the efficiency of H$_2$ formation.  The
product $T^{1/2} y_F(T)$ is likely to be constant to within a factor of 2
\citep{hol71}.

Equation \ref{eqn:nhi} also contains a dependence upon the level of
obscuration in the immediate vicinity of the FUV source.  While the
variable $\chi$ represents the intrinsic FUV flux associated with the
star-forming region, we generally observe an attenuated FUV
flux. Assuming any extinction associated with the star-forming region
is in the form of an overlying screen of optical depth $\tau (FUV)$,
$\chi({\rm observed})=\chi e^{-\tau (FUV)}$.  Since the ISM in the
immediate vicinity of the FUV source will have been disturbed by
stellar winds and any prior supernovae, we do not attempt to link
$\tau (FUV)$, the local obscuration of the FUV source at 1500 \AA, to
that implied by $\delta=A_V/N_H$, the larger-scale dust--to--gas
ratio.

Assuming solar neighborhood values of $\sigma_0=1.883 \times 10^{-21}$
cm$^2$, $D=5.43 \times 10^{-11}$ s$^{-1}$, $R_0=3 \times 10^{-17}$
cm$^{3}$ s$^{-1}$, and $G_0 \approx 5 \times 10^{-5}$
\citep{ste88}, and neglecting the weak
temperature dependence of $R$, equation \ref{eqn:nhi} becomes:
\begin{equation}
N(\HI)= {{5 \times 10^{20}} \over {(\delta/\delta_0)}}
\ln{\left[ {{90 \chi } \over {n}}
{{\left( {{\delta} \over {\delta_0}} \right)}^{-1/2}}
+ 1 \right]}, 
\label{eqn:nhidel}
\end{equation}
where $\chi=\chi({\rm observed})e^{\tau (FUV)}$.

The behavior of $N(\HI)$ as a function of $\chi$ is displayed for
$\delta/\delta_0=0.2$, $\tau (FUV) = 0$ and $n=30$, $n=300$, and
$n=3000$ in Figure \ref{fig:nhichi}.  (These values of $\delta/\delta_0$ 
and $\tau (FUV)$ are appropriate for the outer regions of M101, as 
discussed in Section \ref{sec:volume}.) Values of $N(\HI) \gtrsim
10^{22}$ cm$^{-2}$ are not likely to be observed as the atomic gas
probably becomes optically thick at this point:
\begin{eqnarray}
N(\HI)	& = & 1.82 \times 10^{18} \int_{- \infty}^\infty T_s \tau (v)
dv \nonumber \\
 & \rightarrow & 1.82 \times 10^{18} T_s \tau \Delta v \nonumber \\
 & \approx & 10^{22} {\rm cm}^{-2}, \nonumber
\end{eqnarray}
for spin temperatures of $T_s \approx 100$K and profile FWHMs of
$\Delta v \approx 20$ km s$^{-1}$ typical of M101 \citep{bra97}, and for
optical depths of $\tau \approx 2.5$, corresponding to a ratio between
the brightness and kinetic temperatures of $T_B/T_K \approx 0.9$.
This value is appropriate for the highest-brightness regions of M101, as
indicated in Figure 8a of \citet{bra97}.  Figure \ref{fig:nhichi} also
shows the measurements for each of the 35 candidate PDRs.  The data
indicate that the properties of observed regions in M101 are
consistent with photodissociation of an underlying molecular gas of
moderate volume density.

\subsection{The Volume Density}
\label{sec:volume}

Equations \ref{eqn:nhi} and \ref{eqn:nhidel}
also suggest that our measurements of the \HI\ column
density and the FUV emission may be used to estimate the molecular gas
density in M101.  Solving for the volume density of the gas yields:

\begin{equation}
n   =  {{90 \chi}{\left(\delta \over \delta_0 \right)^{-1/2}} \left[
e^{N_{{\rm HI}}\left(\delta / \delta_0 \right)/5\times10^{20}} -1 \right]^{-1}} 
\label{eqn:voldens}
\end{equation}
for $N_{{\rm HI}} \equiv N(\HI)$ and $\chi=\chi({\rm observed})e^{\tau (FUV)}$.

This method provides a new probe of the molecular gas volume density
with several advantages, including: 1) the method is independent of
the CO/${\rm H_2}$ conversion factor, 2) the FUV emission can be
directly measured from the UIT data, 3) the 21 cm line emission is
insensitive to the excitation conditions in the ISM.  The main
disadvantages are: 1) the results are strongly dependent on the value
of $\delta = A_V/N_H$, which will vary throughout the galaxy, 2) the
relationships contain a weak dependence on the gas temperature which
is not yet well understood, 3) the amount of extinction affecting the 
FUV fluxes must be evaluated, and 4) the observed values of N(\HI)
represent upper limits since the unknown geometry of the individual
PDRs prohibits inclination corrections.  This method is also biased
towards regions where significant amounts of gas are present,
i.e.~regions which are massive enough to have formed some dozens of
O--B stars in the last 10--100 $\times 10^6$ years. 

\subsubsection{The Dust--to--Gas Ratio}

Values of $n$ derived from the data in Table \ref{tab:data}, assuming
a constant Milky Way dust--to--gas ratio $\delta/\delta_0=1$ and no
extinction ($\tau(FUV)=0$) are also given in Table \ref{tab:data} as
$n_{raw}$.  These values suggest that the volume density of the gas
may decrease as a function of radius.  However, the metallicity in
M101 also changes strongly with radius (e.g.~\citet{ken96} and
references therein).  Since metal--rich regions also tend to be
dustier, the observed radial trend in volume density most likely
reflects a change in the dust--to--gas ratio.

To investigate this possibility, a measure of the dust--to--gas ratio
is needed.  Following \citet{iss90} and \citet{sch93}, we assume that
$\delta=A_V/N_H$ is directly proportional to the metallicity, as
traced by the oxygen abundance.  Radial profiles of the quantity $[12
+ (\log {\rm O/H})]$ for M101 are given in \citet{ken96} for different
values of the calibration between the $R_{23}=$ ([O\II ] $+$ [O \III
]$)/{\rm H}\beta$ abundance parameter and oxygen abundance.  The radial
profiles appear to be linear for the \citet{dop86} and \citet{mcc85}
calibrations or to steepen somewhat in the central regions for the
\citet{edm84} calibration.  In the rest of the paper, we refer to the
metallicity gradients in terms of the calibration source, but remind
the reader that the gradients all refer to the \citet{ken96} data.
Figure \ref{fig:dustgas} shows the range of radial dependence of the
dust--to--gas ratio for M101.  Values of $n$ derived using
dust--to--gas ratios based upon the Edmunds \& Pagel type metallicity
gradient are shown in Figure \ref{fig:volcorr}.  This figure shows
values increasing from $n \sim 10$ cm$^{-3}$ in the central star
forming regions of M101 to $n \sim 1000$ cm$^{-3}$ in the periphery of
the galaxy. The trend is similar for the Dopita \& Evans or the MRS
type metallicity gradients, although slightly lower values are
obtained ($n \sim 0.1$ cm$^{-3}$ to $n \sim 100$ cm$^{-3}$).  Values 
of $n$ derived using an average of the Dopita \& Evans and MRS type 
metallicity gradients are shown in Figure \ref{fig:volcorrb}.

\subsubsection{Obscuration Effects} 
\label{sec:tau}

The discussion up to this point has assumed that extinction effects in
these FUV--bright regions are negligible, based upon the fact that the
clear majority of H$\alpha$ sources have FUV counterparts.  Studies of
FUV--bright regions in M81 and M51 suggest that such regions are still
obscured by dust with optical depths ranging from $\tau_V \sim 0.4$ to
$\tau_V \sim 3.2$ \citep{hil95,hil97}, however.  These values would
translate to non--negligible reddenings of at least 1 to 9 magnitudes
at 1500 \AA.  We use the H$\alpha$ and H$\beta$ measurements of
\citet{sco92} to apply an extinction correction to our values of
$\chi$.  These authors find that the central regions of M101 are
characterized by higher optical depths than the outer regions,
consistent with the general behavior of the dust--to--gas ratio.
Based upon the scatter plot of optical depths in Figure 3 of Scowen et
al., we adopt a radial profile of the optical depth of the form
$\tau_V=2.3-R/6.5$, where $R$ is the radius, for $R \le 15$ kpc, and
$\tau_V=0$ for $R > 15$ kpc.  These values provide a lower limit to the
required extinction corrections, as radio observations of \HII\
regions in M101 indicate that optical depths derived from optical
H$\alpha$/H$\beta$ line ratios only trace a fraction of the total
optical depth in those regions \citep{via82}.  The ensuing
extinction--corrected values of $n$ are shown in Figure \ref{fig:tau1}
for the \citet{sav79} extinction curve ($\tau (FUV)=2.6 \tau_V$) and
\citet{edm84} type metallicity gradient.  Even higher volume densities
(approaching $n \sim 2000$ cm$^{-3}$ in the central regions) are
obtained with the \citet{kin94} extinction curve.  Figure
\ref{fig:tau2} shows extinction--corrected values of $n$ for the
\citet{sav79} extinction curve and the average of the Dopita \& Evans
and MRS type metallicity gradients.

These calculations use corrections based upon averaged
extinction and dust--to--gas values; the volume densities could be
further constrained by obtaining data on the specific optical depths
and dust--to--gas ratios of the specific 35 candidate PDRs in our
study.  

\section{Discussion and Conclusions} \label{sec:imp}

Figures \ref{fig:tau1} and \ref{fig:tau2} are our current best
estimates for the radial dependence of the total gas volume density in
M101. This gas should be essentially all \Htwo. In the process of
correcting the ``raw'' values of volume density (Table 
\ref{tab:data}) for the radial variations of dust-to-gas fraction
(resulting in Figures \ref{fig:volcorr} and \ref{fig:volcorrb}) and UV
extinction (finally producing Figures \ref{fig:tau1} and
\ref{fig:tau2}), the radial variations have disappeared. The molecular
gas shows a surprisingly narrow range of values from 30 - 1000 \Htwo\
molecules cm$^{-3}$ for the Edmunds \& Pagel type metallicity gradient,
with no clear trend from the inner \HI\ --deficient regions near $R_G
\lesssim 5$ kpc, through the \HI\ --rich ``main body'' of the galaxy
near $R_G \approx 15$ kpc, all the way out to the \HI\ --poor outer
parts at $R_G \gtrsim 25$ kpc.  We note that the Edmunds \& Pagel 
type metallicity gradient is a good representation for the mean of an 
ensemble of nearby galaxies \citep{pan00}, which leads us to 
favor Figure \ref{fig:tau1} as the most likely final result.

One concern is that the narrow range of values in Figure
\ref{fig:tau1} could result from observational selection.  For
example, regions of high volume density located near typical FUV
sources may be missed since the \HI\ would have a relatively low
column density and a low filling factor.  This is not an issue for
this study since only 3\% of our initial sample of FUV sources were
rejected due to a lack of 21 cm line emission.  To confirm this
statement, we examined an additional 10 distinct FUV
sources. Measurable \HI\ is found within 800 pc for all ten 
sources.  This is consistent with the spatial correlation between FUV
sources and 21 cm line emission observed in Figures
\ref{fig:overlay2}, \ref{fig:fuvhi2}, and \ref{fig:fuvhi3}.  Virtually
all FUV sources are associated with 21 cm line emission, with the
exception of the innermost regions of M101, where the \HI\ column
density is supressed due to the strong increase in the dust-to-gas
ratio.

What kind of observational selection could operate on the low-density
side of Figure \ref{fig:tau1}?  According to Equation \ref{eqn:nhi},
\Htwo\ with low volume density can produce large columns of \HI\ which
may become optically thick, leading us by Equation \ref{eqn:voldens}
to overestimate the \Htwo\ density and thus depopulate the lower
part of Figure \ref{fig:tau1}.  However, gas of density e.g.~$n \lesssim
10$ cm$^{-3}$ with \HI\ columns in the optically-thick regime of
N(\HI) $\approx 10^{22}$ cm$^{-2}$ would be typically $\gtrsim 300$ pc
in size.  Examination of Figure \ref{fig:overlay2} and the zoomed
Figures \ref{fig:fuvhi2} and \ref{fig:fuvhi3} shows that the
highest-brightness \HI\ regions are unresolved, however. NGC~5447 
in Figure \ref{fig:fuvhi3} is a good example of this, as are 
NGC~5461 and NGC~5462 in Figure \ref{fig:fuvhi2}.  This means 
the path lengths are $\lesssim 200$ pc, which implies values 
of N(\HI) $\lesssim 6 \times 10^{21}$ cm$^{-2}$ in Figure \ref{fig:nhichi}, 
where optical depth in the \HI\ is not likely to be a problem. 
Therefore, the paucity of points at low volume densities in 
Figure \ref{fig:tau1} is very likely real.  

We conclude that the 35 regions studied do indeed comprise a
representative sample of PDRs, and that observational selection 
is not likely to be the cause of the narrow range of values 
for $n$(\Htwo) in Figure \ref{fig:tau1}.

Our results refer only to the ISM in the immediate vicinities of young
stars and clusters of young stars. We do not know from the present
observations if the results are representative of the ISM as a
whole. Furthermore, the measurement of the \HI\ emission does not
immediately provide a total gas mass without additional assumptions
about the geometry of the ISM since the \HI\ emission is a surface
phenomenon in the PDR picture.  Further discussion of the implications
of our results for the ISM content of galaxies in general is beyond
the scope of this paper; we hope to return to these questions in a
future publication.

\citet{wal96} have proposed an interpretation for the radial
distribution of the bright \HI\ peaks in M101 and other nearby spirals
in terms of variations in the hydrostatic pressure of the ISM. Their
picture is insightful and deserves a more thorough discussion
in the light of the present results than space here permits. 

UV photons clearly dominate the physics of the
ISM in the immediate neighborhoods of young stars. The production of
\HI\ from \Htwo\ by photodissociation is a natural and inevitable
consequence of this physics. We have described how photodissociation
can explain the geometrical structure of the \HI\ on scales of $\sim
100$ pc in nearby galaxies and have discussed how the physics of
photodissociation can be used to provide a new probe of the density of
the underlying molecular hydrogen. In addition, we have presented a 
straightforward explanation for the disappearance of \HI\ in the 
inner parts of galaxies. Acceptance of this picture requires
a shift in parts of the paradigm for star formation from the ISM, from
viewing \HI\ as a precursor, to seeing it as a \textit {product} of the
star formation process. While the present results do not by themselves
demand discarding the old paradigm, they further strengthen the
viability of the new one.

\acknowledgements

Portions of this reseach were funded by NASA grant NAG5-1278 to the
STScI. The UIT was funded under NASA Project Number 440-51.  We wish
to thank L.~van Zee, M.~Haynes, \& R.~Braun for providing the
H$\alpha$ and \HI\ data, and P.~van der Kruit for comments on an
earlier version of this paper. We are grateful to our referee for a
careful reading of our manuscript and comments which improved the 
paper. The authors have made use of the NASA/IPAC Extragalactic
Database (NED) which is operated by the Jet Propulsion Laboratory,
Caltech, under contract with the National Aeronautics and Space
Administration.


\onecolumn 

\clearpage

\begin{figure}
\figcaption[paperfuvha_new.pps]{FUV and H$\alpha$ emission. 
Contours of the FUV emission are superposed on a greyscale
representation of the H$\alpha$ emission. The contour levels 
are $1.5 \times 10^{-17}$, $5 \times 10^{-17}$, $1 \times 10^{-16}$, 
$5 \times 10^{-16}$, and $1 \times 10^{-15}$ \UITunits.  Boxes 1 and 2 
indicate the regions shown in more detail in Figures \protect\ref{fig:fuvha2} 
and \protect\ref{fig:fuvha3}.
\label{fig:overlay1}}
\end{figure}

\begin{figure}
\figcaption[posterfuvhi_new.pps]{FUV and 21 cm line emission. Contours of
the FUV emission are superposed on a greyscale representation of the
21 cm line emission. Contour levels and boxes are as in Figure
\protect\ref{fig:overlay1}. \label{fig:overlay2}}
\end{figure}

\begin{figure}
\epsscale{0.8}
\figcaption[paperfuvhasec2_fin.pps]{FUV and H$\alpha$ emission -
region 1. The detailed behavior of the H$\alpha$ emission from region
1 (see Figure \protect\ref{fig:overlay1}) is shown in grayscale.  Contours of
the FUV emission are given for FUV fluxes of $1.5 \times
10^{-15}$, $4 \times 10^{-17}$, $6 \times 10^{-17}$, $8 \times
10^{-17}$, $1 \times 10^{-16}$, $1.5 \times 10^{-16}$, $2 \times
10^{-16}$, $3 \times 10^{-16}$, and $5 \times 10^{-16}$ \UITunits.
H$\alpha$ sources without FUV emission are labelled according to 
their Hodge numbers. 
\label{fig:fuvha2}}
\end{figure}

\begin{figure}
\epsscale{0.8}
\figcaption[paperfuvhisec2_fin.pps]{FUV continuum and 21 cm line emission -
region 1.  The detailed behavior of the 21 cm line emission from
region 1 (see Figure \protect\ref{fig:overlay2}) is shown in grayscale.
Contours of the FUV emission are given as in Figure
\protect\ref{fig:fuvha2}.  The FUV contours appear smooth in
comparison to the 21 cm line emission since the ultraviolet data were
not resampled after smoothing; the FUV emission is essentially
oversampled in this representation. \label{fig:fuvhi2}}
\end{figure}

\begin{figure}
\epsscale{0.8}
\figcaption[paperfuvhasec3_fin.pps]{FUV and H$\alpha$ emission - region 2. 
The detailed behavior of the H$\alpha$ emission from region
2 (see Figure \protect\ref{fig:overlay1}) is shown in grayscale.  Contours of
the FUV emission are given for FUV fluxes of $1.5 \times
10^{-15}$, $4 \times 10^{-17}$, $6 \times 10^{-17}$, $8 \times
10^{-17}$, $1 \times 10^{-16}$, $1.5 \times 10^{-16}$, $2 \times
10^{-16}$, and $5 \times 10^{-16}$ \UITunits.  Sources A, B, and 
C are FUV sources lacking significant H$\alpha$ emission. 
\label{fig:fuvha3}}
\end{figure}

\begin{figure}
\epsscale{0.8}
\figcaption[paperfuvhisec3_fin.pps]{FUV continuum and 21 cm line emission - region 2. 
The detailed behavior of the 21 cm line emission from
region 2 (see Figure \protect\ref{fig:overlay2}) is shown in grayscale.
Contours of the FUV emission are given as in Figure
\protect\ref{fig:fuvha3}.  
\label{fig:fuvhi3}}
\end{figure}

\begin{figure}
\figcaption[paperfuv.pps]{FUV continuum emission.  The FUV image
shown here has been smoothed and regridded to match the spatial
resolution and pixel spacing of the \HI\ data.  The locations of 35
relatively isolated sources are shown. Squares correspond to regions
with $0.9 < \chi < 10$.  Regions with $\chi$ values outside of this
range are marked by circles. \label{fig:fuv}}
\end{figure}

\clearpage 

\clearpage
\begin{figure}
\epsscale{0.8}
\plotone{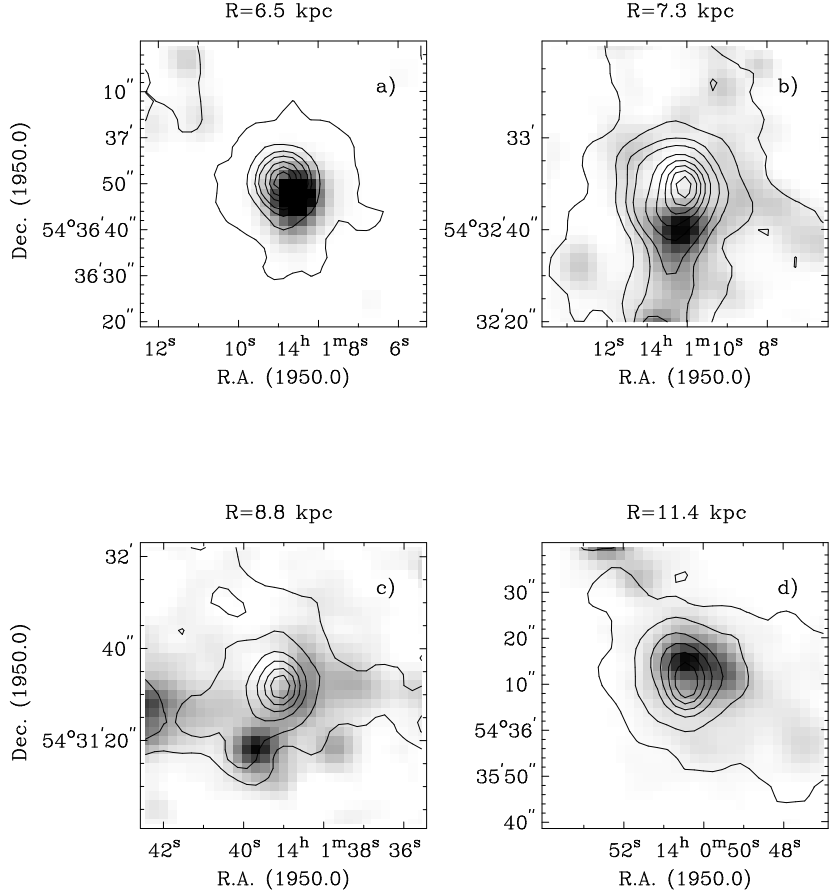}
\figcaption[paperpdrha1_new.pps,paperpdrha2_new.pps]{PDR Morphology: FUV and
H$\alpha$.  This figure shows the FUV and H$\alpha$ emission from 
eight of the candidate PDRs with $0.9 <
\chi < 10$.  The regions are in order of increasing distance from the
nucleus of M101.  The individual values of $\chi$
are a) 5.8, b) 3.0, c) 3.9, d) 2.0, e) 2.3, f) 2.5, g) 7.0, and h)
1.8.  The FUV contour levels are, in units of $1 \times 10^{-16}$ 
\UITunits, a) 0.25, 0.5, 0.75, 1, 1.5, 2, 2.5, and 3, 
b) 0.25, 0.5, 0.75, 1, 1.5, 2, 2.5, 3, and 3.5, 
c) 0.4, 0.6, 1, 1.4, 1.8, 2.2, and 2.6,
d) 0.25, 0.5, 1, 1.5, 2, 2.5, and 3, 
e) 0.2, 0.4, 0.8, 1.2, 1.6, and 2, 
f) 0.2, 0.4, 0.8, 1.2, and 1.6,
g) 0.25, 0.5, 1, 2, 3, 4, and 5, and 
h) 0.10, 0.20, 0.30, and 0.40.
\label{fig:pdrfuvha}}
\end{figure}

\clearpage
\begin{figure}
\figurenum{8}
\epsscale{0.8}
\plotone{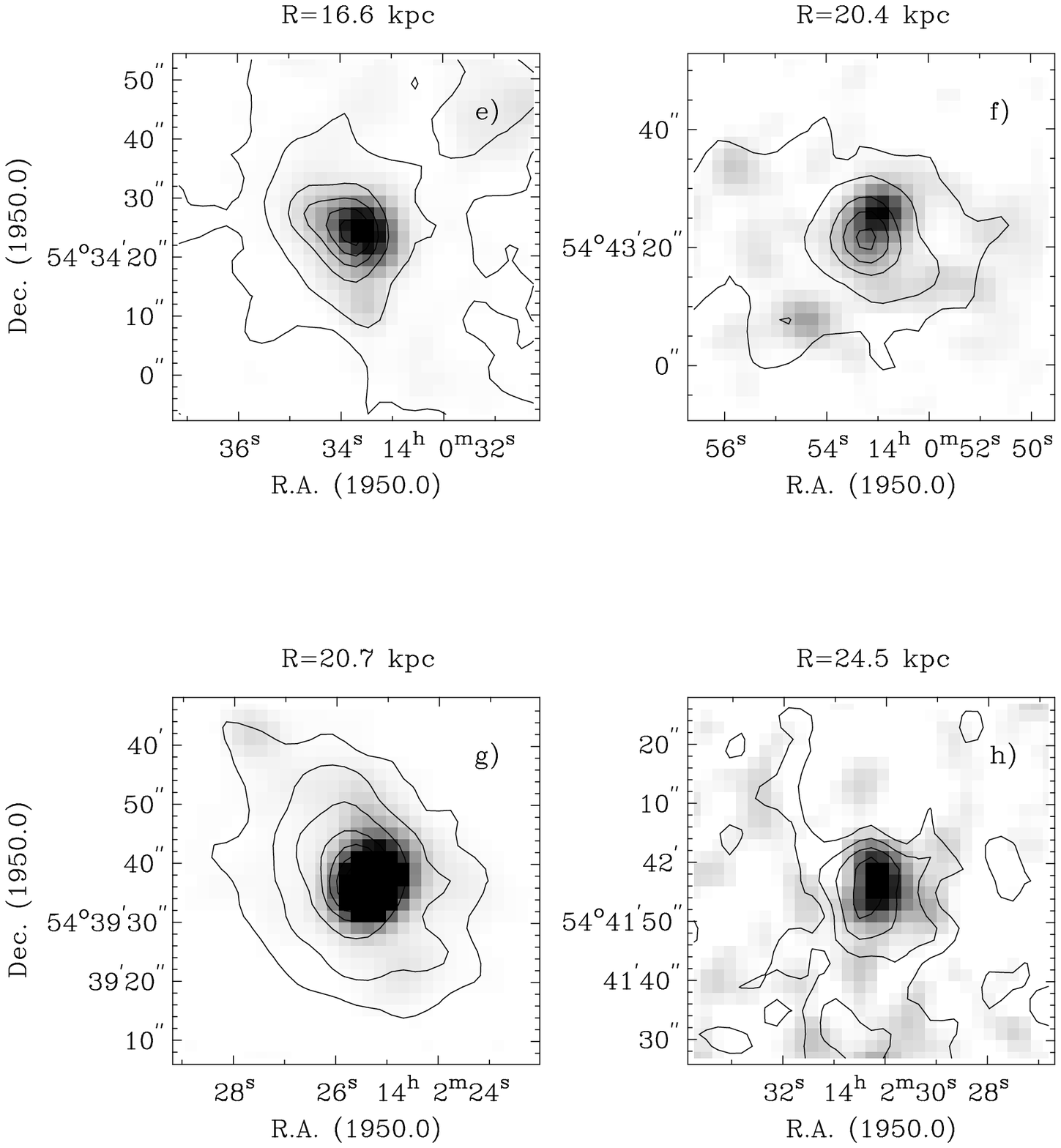}
\figcaption[paperpdrha1_new.pps,paperpdrha2_new.pps]{PDR Morphology: FUV and
H$\alpha$, continued.  }
\end{figure}

\clearpage
\begin{figure}
\epsscale{0.8}
\plotone{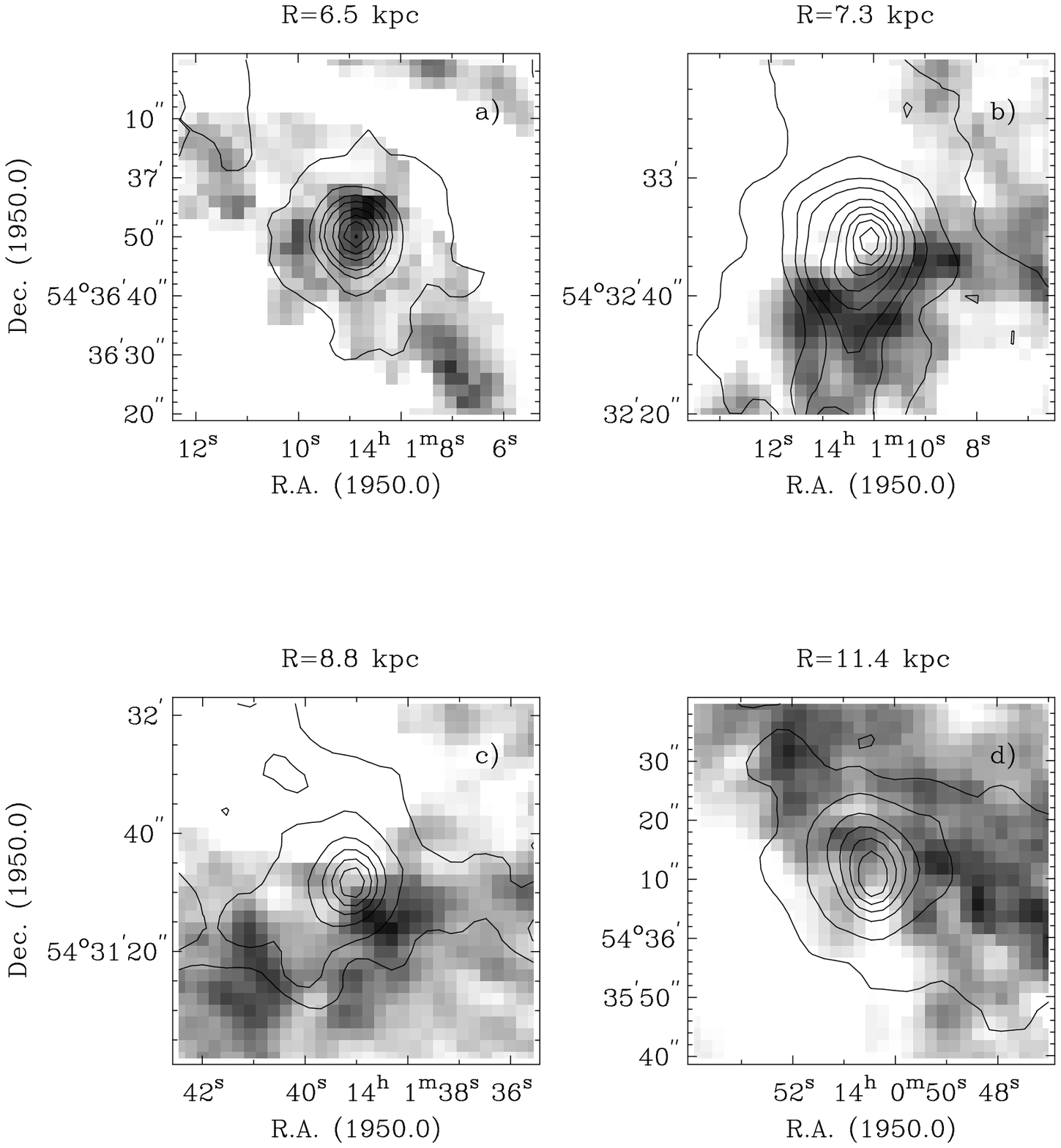}
\figcaption[paperpdrhi1.pps,paperpdrhi2.pps]{PDR Morphology: FUV contours and
\HI\ grayscale.  As in Figure \protect\ref{fig:pdrfuvha}, for the FUV continuum
and 21 cm line emission. The \HI\ is offset from and partially
surrounds the FUV source.  The area filling factor appears to be
roughly $\sim 30$\%, independent of radial distance. \label{fig:pdra}}
\end{figure}

\clearpage
\begin{figure}
\figurenum{9}
\epsscale{0.8}
\plotone{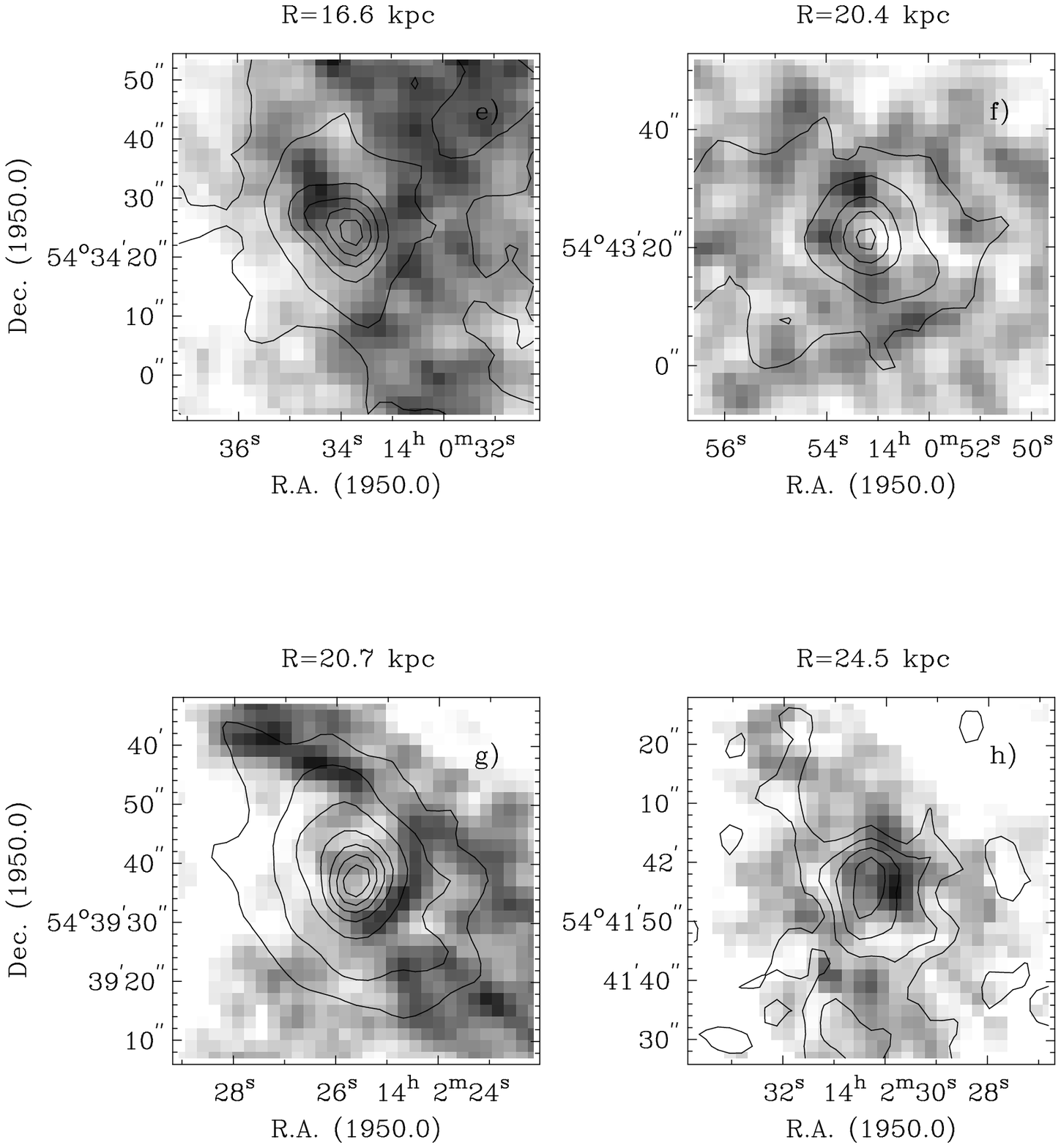}
\figcaption[paperpdrhi1.pps,paperpdrhi2.pps]{PDR Morphology: FUV contours and
\HI\ grayscale, continued.}
\end{figure}

\clearpage
\begin{figure}
\epsscale{0.9}
\plotone{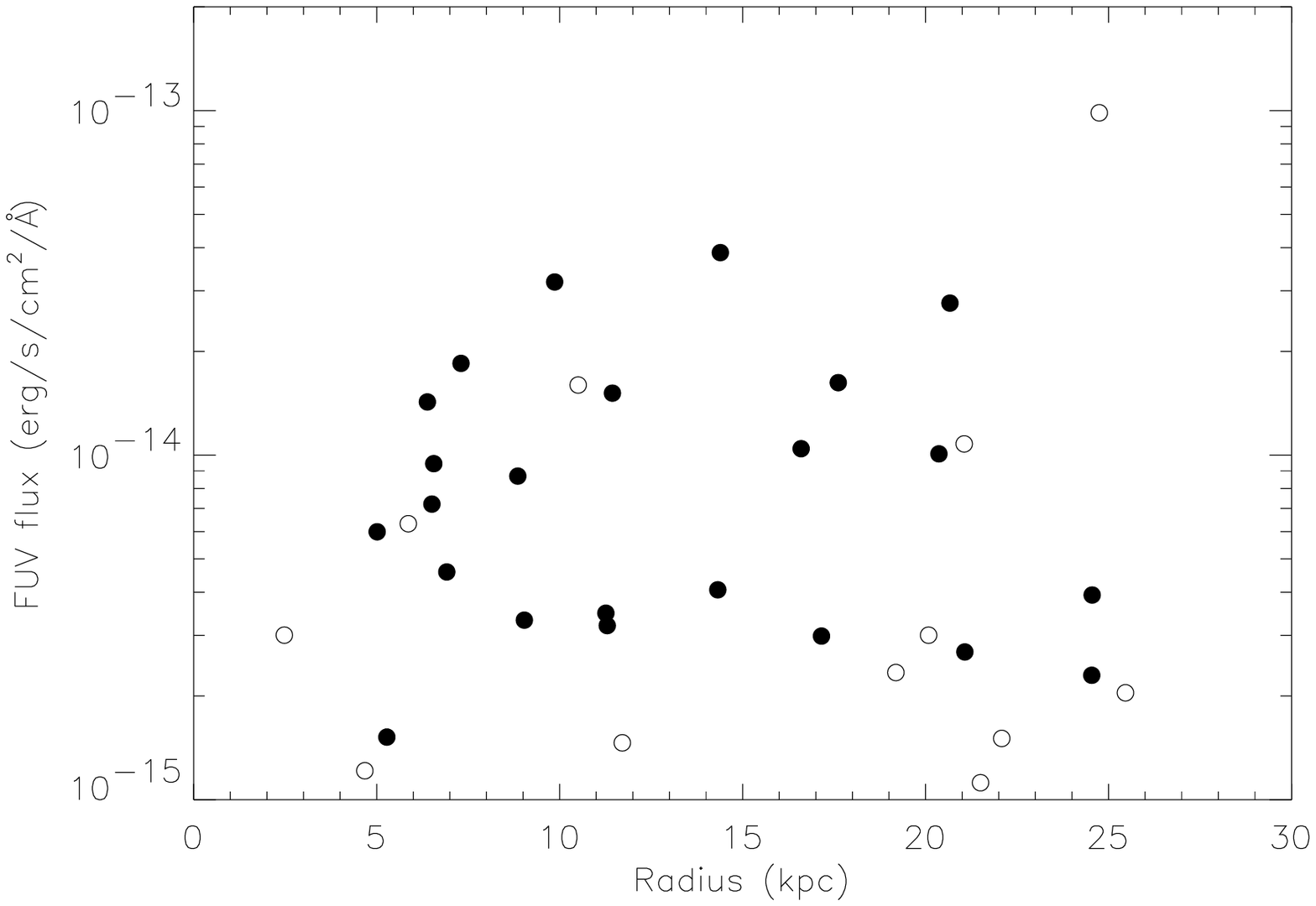}
\figcaption[fuvflux.eps]{Observed FUV fluxes ($F_{FUV}$).  This figure illustrates the range of FUV fluxes produced by the 35 isolated sources, as observed by UIT. Diffuse FUV emission, as well as any sky emission, have been subtracted.  Filled circles correspond to PDRs with $0.9 < \chi < 10$. PDRs with $\chi$ values outside of this range are denoted by open circles.  The observed FUV fluxes do not show a radial trend. \label{fig:fuvflux}}
\end{figure}

\clearpage
\begin{figure}
\epsscale{0.9}
\plotone{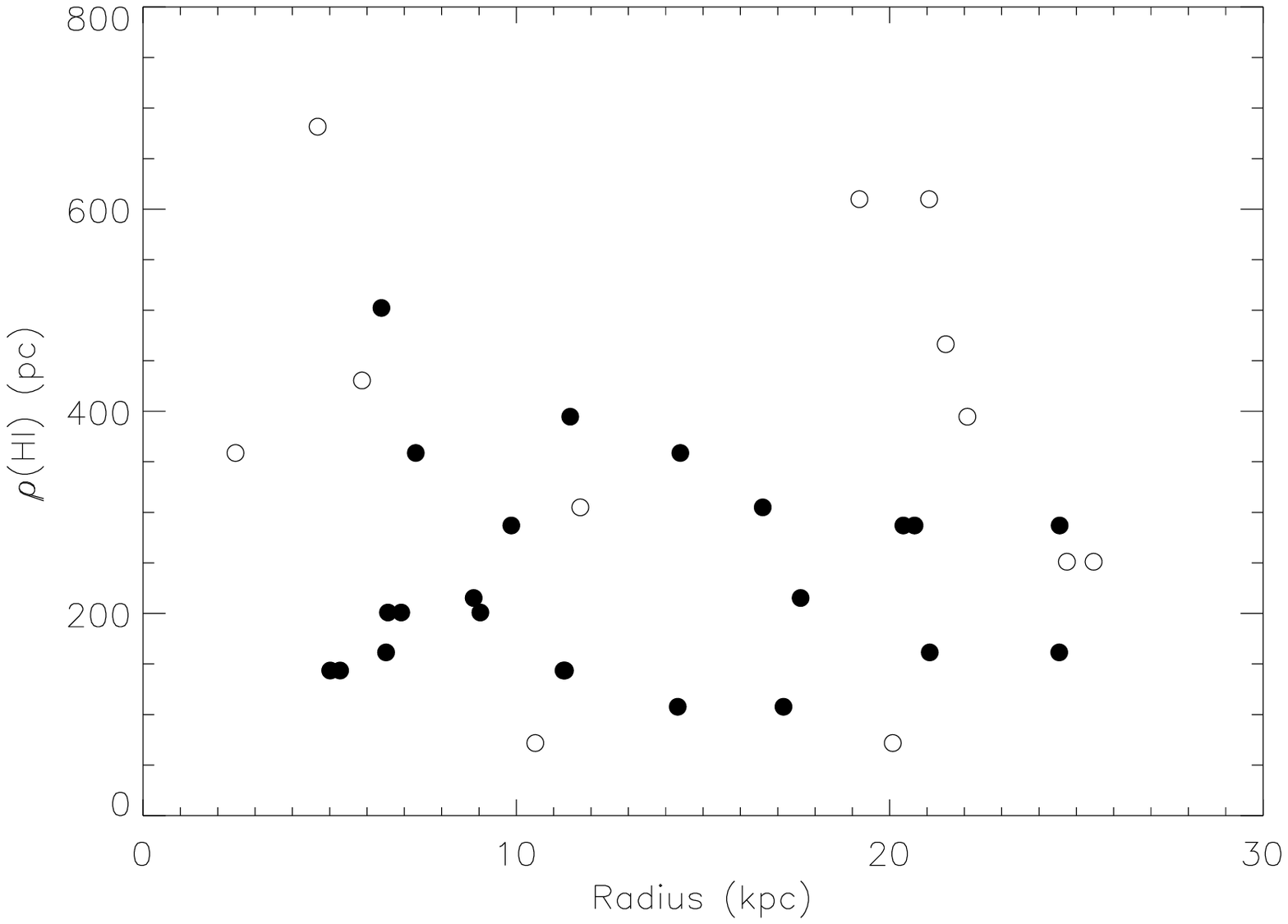}
\figcaption[rhohi.eps]{Observed FUV--\HI\ separation ($\rho_{{\rm HI}}$). The distance between each FUV peak and the peak of the surrounding 21 cm emission is shown.  The quantity $\rho_{{\rm HI}}$ does not show a clear trend with radius. \label{fig:rhohi}}
\end{figure}

\clearpage
\begin{figure}
\epsscale{0.9}
\plotone{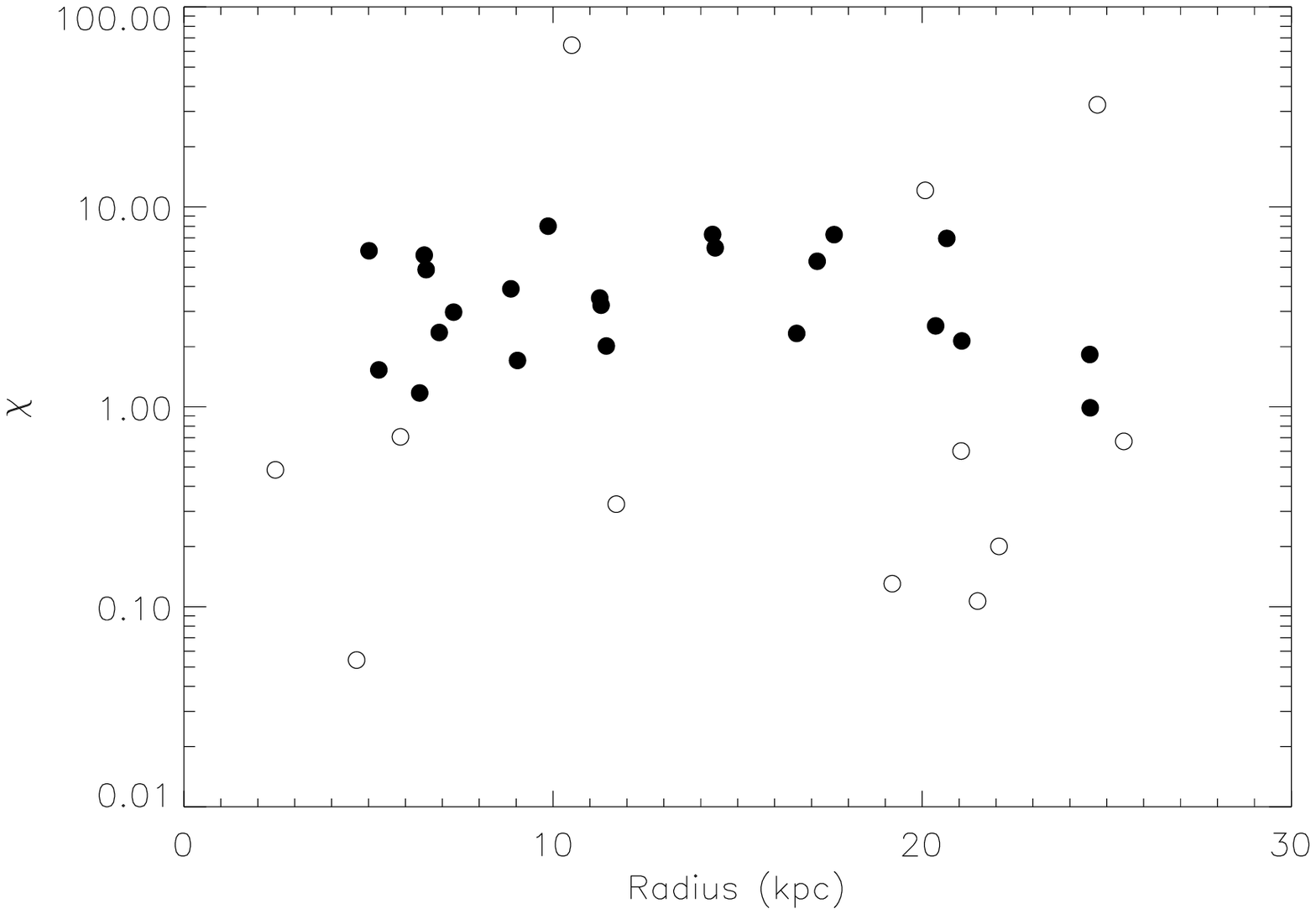}
\figcaption[chi.eps]{Derived $\chi$ Values.  The FUV flux observed at the location of the peak \HI\ is derived from $F_{FUV}$ and $\rho_{{\rm HI}}$.  The values of $\chi$ are independent of radius and clustered between $\chi=0.9$ and $\chi=10.$ \label{fig:chi}}
\end{figure}

\clearpage
\begin{figure}
\epsscale{0.9}
\plotone{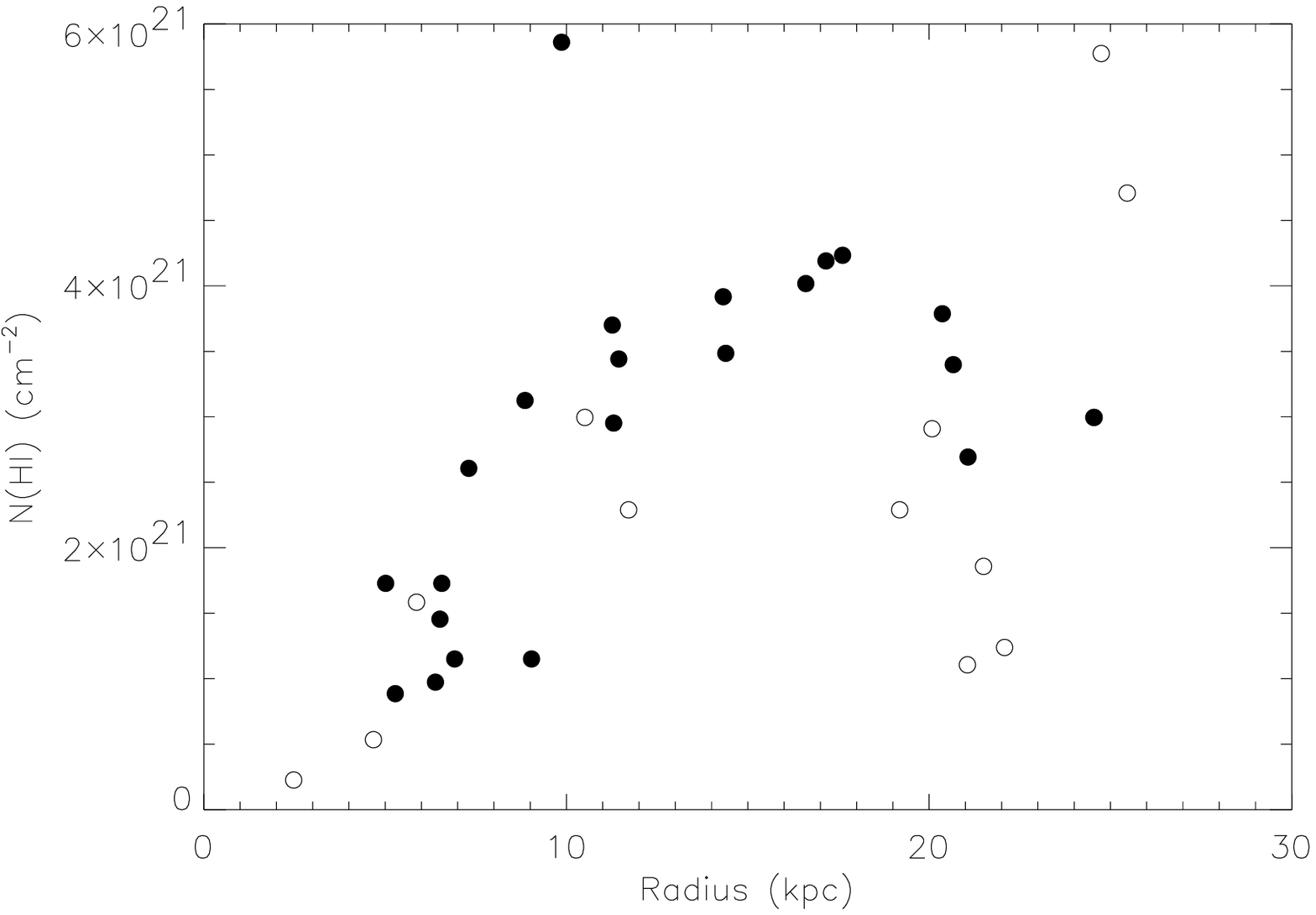}
\figcaption[nhi.eps]{Peak Column Densities ($N_{HI}$).  The peak column density in the vicinity of each FUV source is measured from the \HI\ map.  Column densities generally increase as the distance from the nucleus increases, reflecting the large--scale trend seen in Figure \protect\ref{fig:overlay2}. \label{fig:nhi}}
\end{figure}

\clearpage 

\clearpage
\begin{figure}
\epsscale{0.9}
\plotone{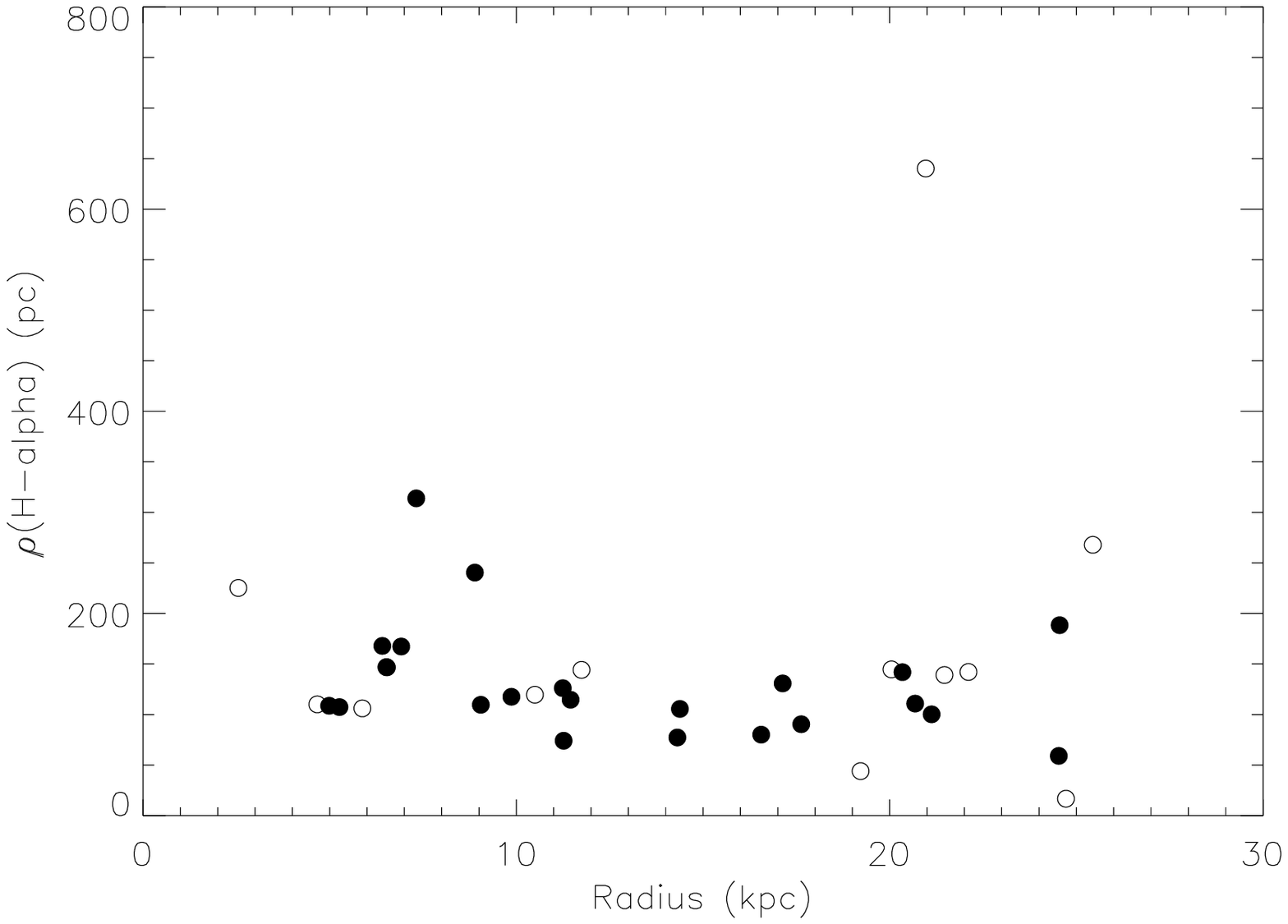}
\figcaption[rhoha.eps]{Observed FUV--H$\alpha$ separation ($\rho_{H\alpha}$). The distance between each FUV peak and the peak of the associated H$\alpha$ emission is shown. \label{fig:rhoha}}
\end{figure}

\clearpage
\begin{figure}
\epsscale{0.9}
\plotone{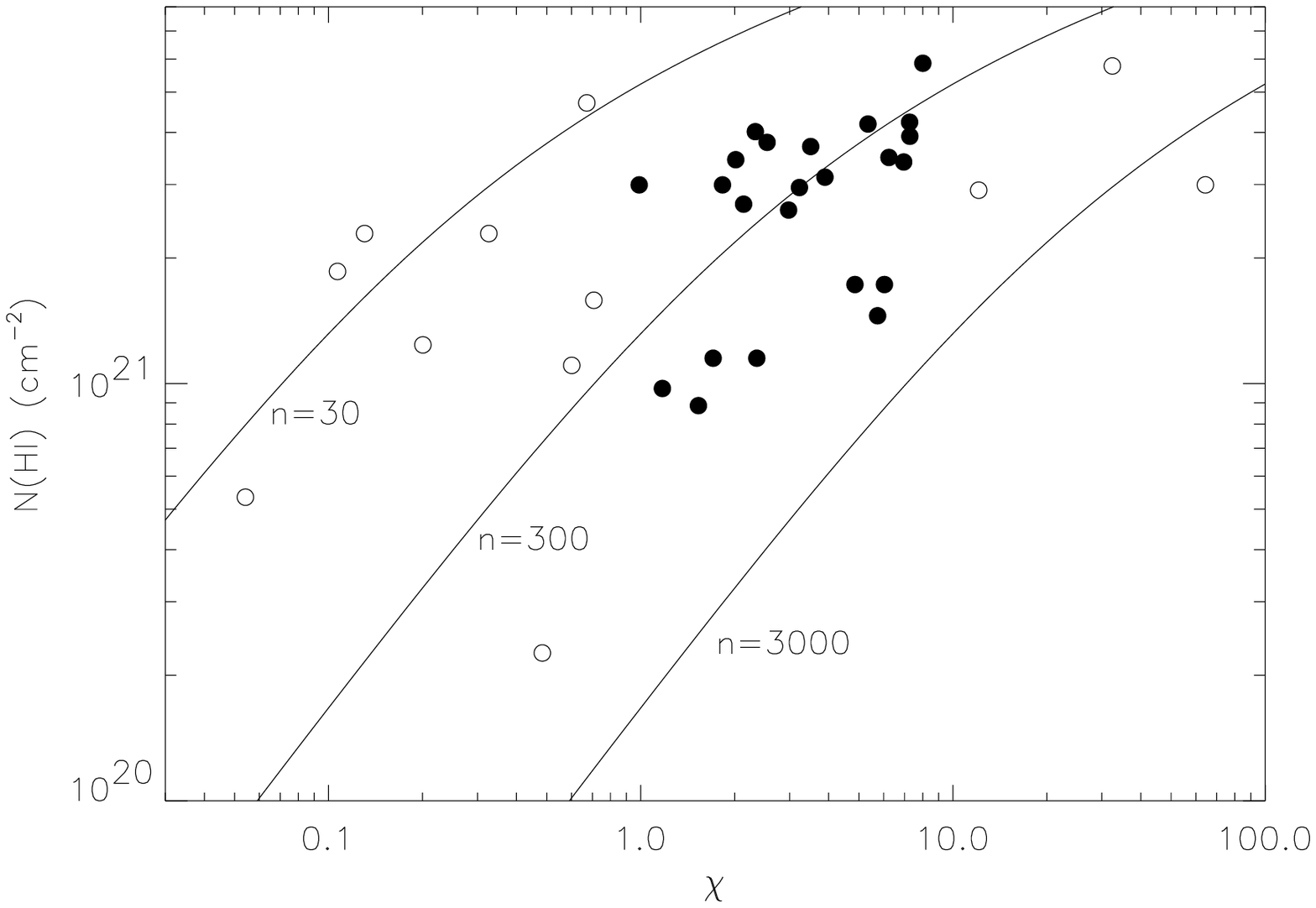}
\figcaption[nhivschimod_final.eps]{The Relationship between $N(\HI)$ and
$\chi$.  The observed and predicted values of $N(\HI)$ are shown as a
function of $\chi$. The modeled behavior of $N(\HI)$ assumes values of
$\delta/\delta_0=0.2$ and $\tau (FUV)=0$, appropriate for the outer regions
of M101.  The observations are clearly consistent with the physics
underlying the photodissociation picture. \label{fig:nhichi}}
\end{figure}

\clearpage
\begin{figure}
\epsscale{0.9}
\plotone{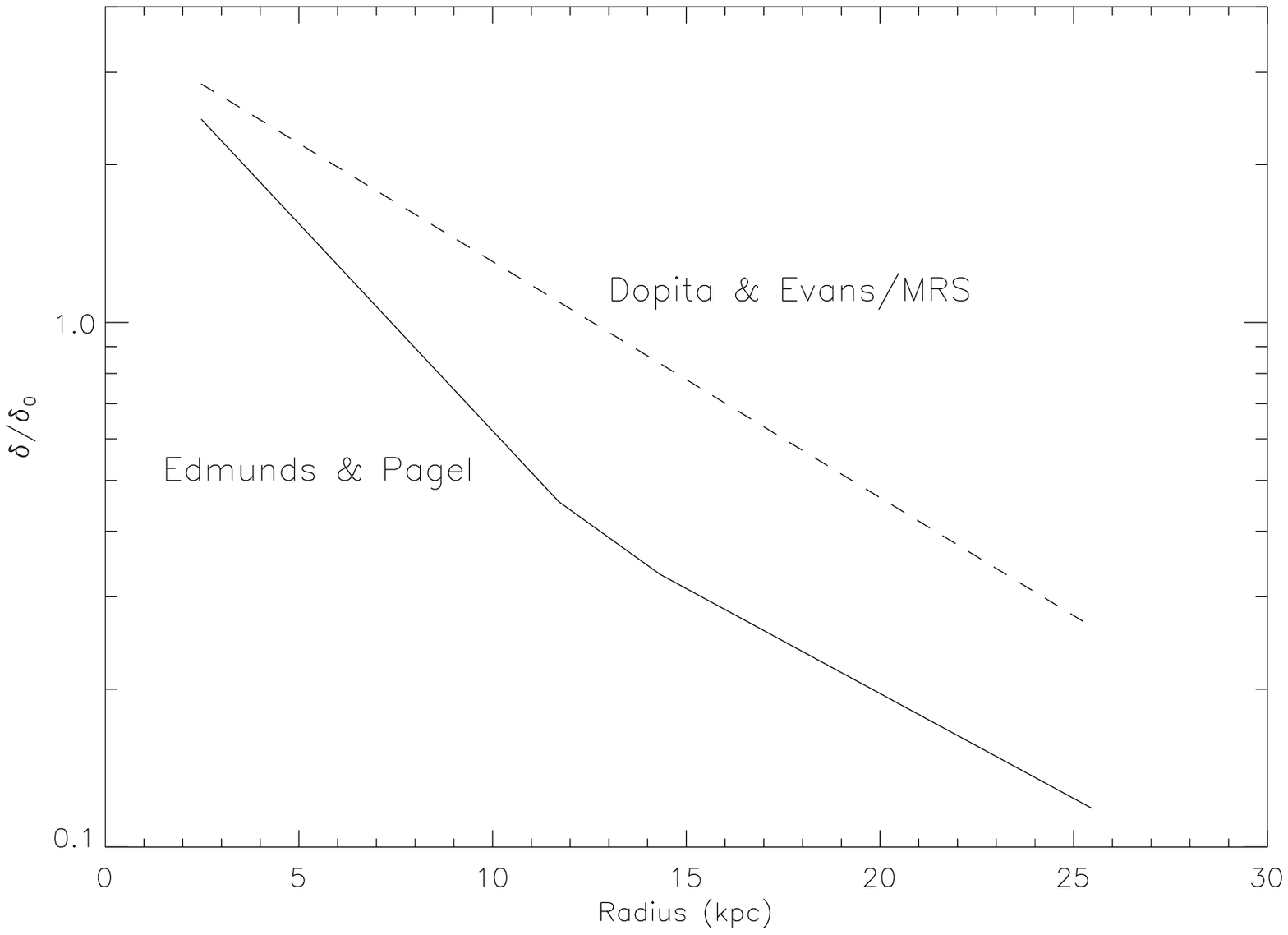}
\figcaption[dustgas.eps]{Dust--to--Gas Profiles.  The illustrated dust--to--gas profiles are obtained from the metallicity gradients given in \protect \citet{ken96}.  The solid line is based upon the calibration of \protect \citet{edm84}; the dashed line reflects the average of the calibrations of \protect\citet{dop86} and MRS. \label{fig:dustgas}}
\end{figure}

\clearpage
\begin{figure}
\epsscale{0.9}
\plotone{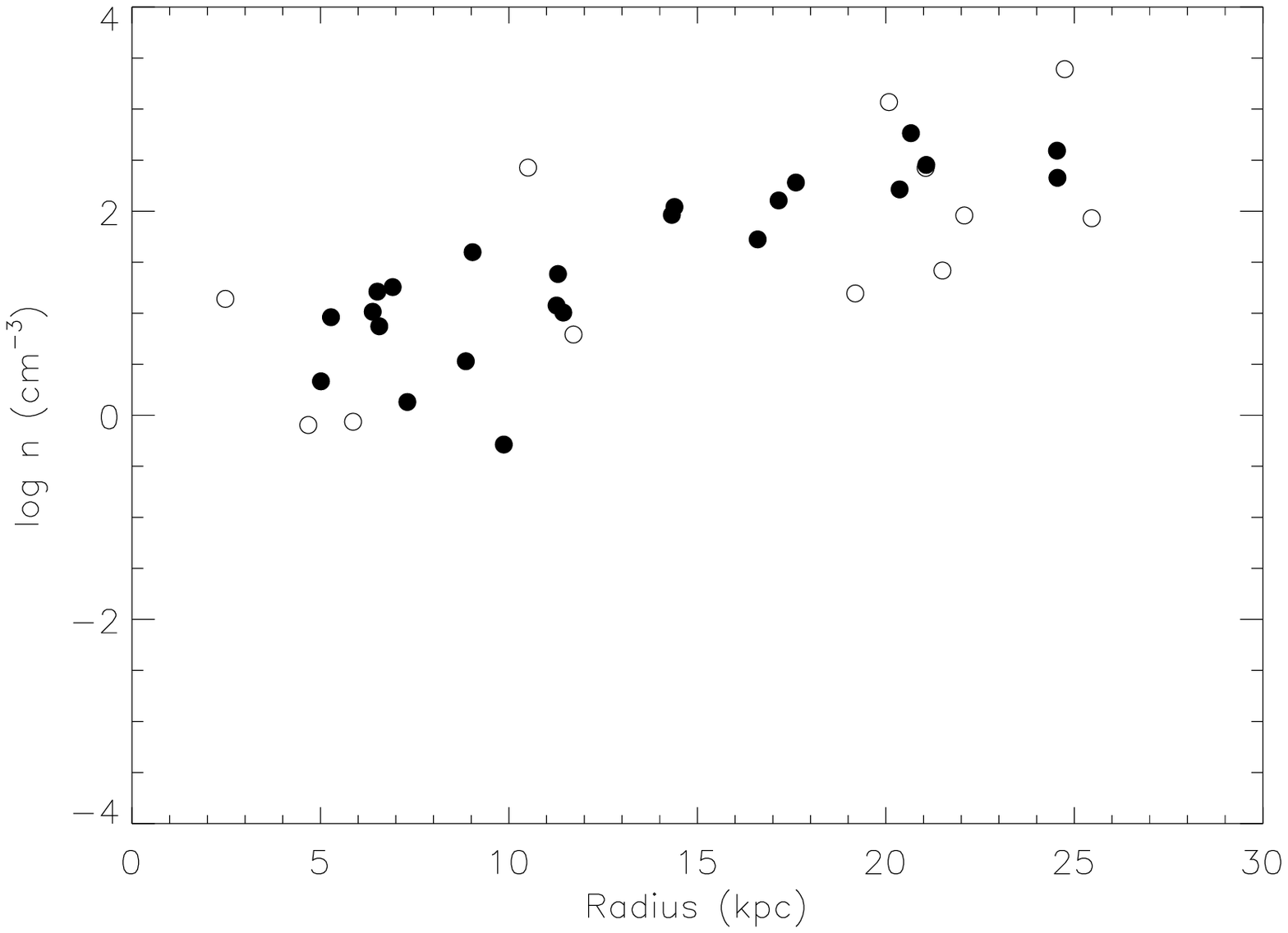}
\figcaption[volZ.eps]{Gas Volume Density with Varying $A_V/N_H$, but no
Extinction Correction.  In the case where the
metallicity gradient steepens in the central regions
(\protect\citet{edm84} calibration), values of $n$ range from $n \sim 10$ cm$^{-3}$
to $n \sim 1000$ cm$^{-3}$. The volume
density appears to increase with radius. $\chi$ has not been corrected 
for extinction. \label{fig:volcorr}}
\end{figure}

\clearpage
\begin{figure}
\epsscale{0.9}
\plotone{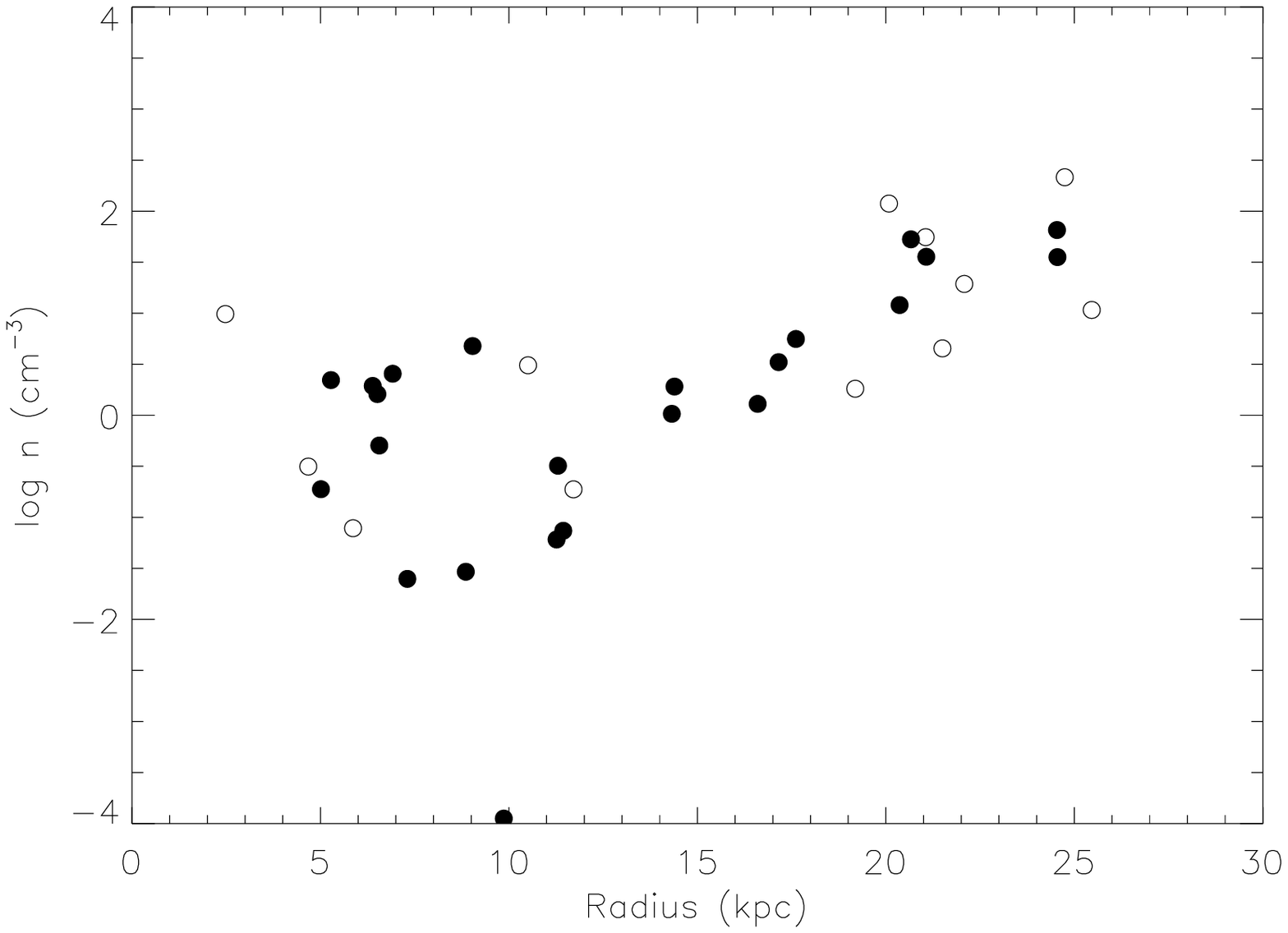}
\figcaption[volZdg.eps]{As in Figure \protect
\ref{fig:volcorr}, but with an alternative metallicity correction
using the average of the calibrations of 
\protect\citet{dop86} and MRS. Note that $\chi$ has not been
corrected for extinction. \label{fig:volcorrb}}
\end{figure}

\clearpage
\begin{figure}
\epsscale{0.9}
\plotone{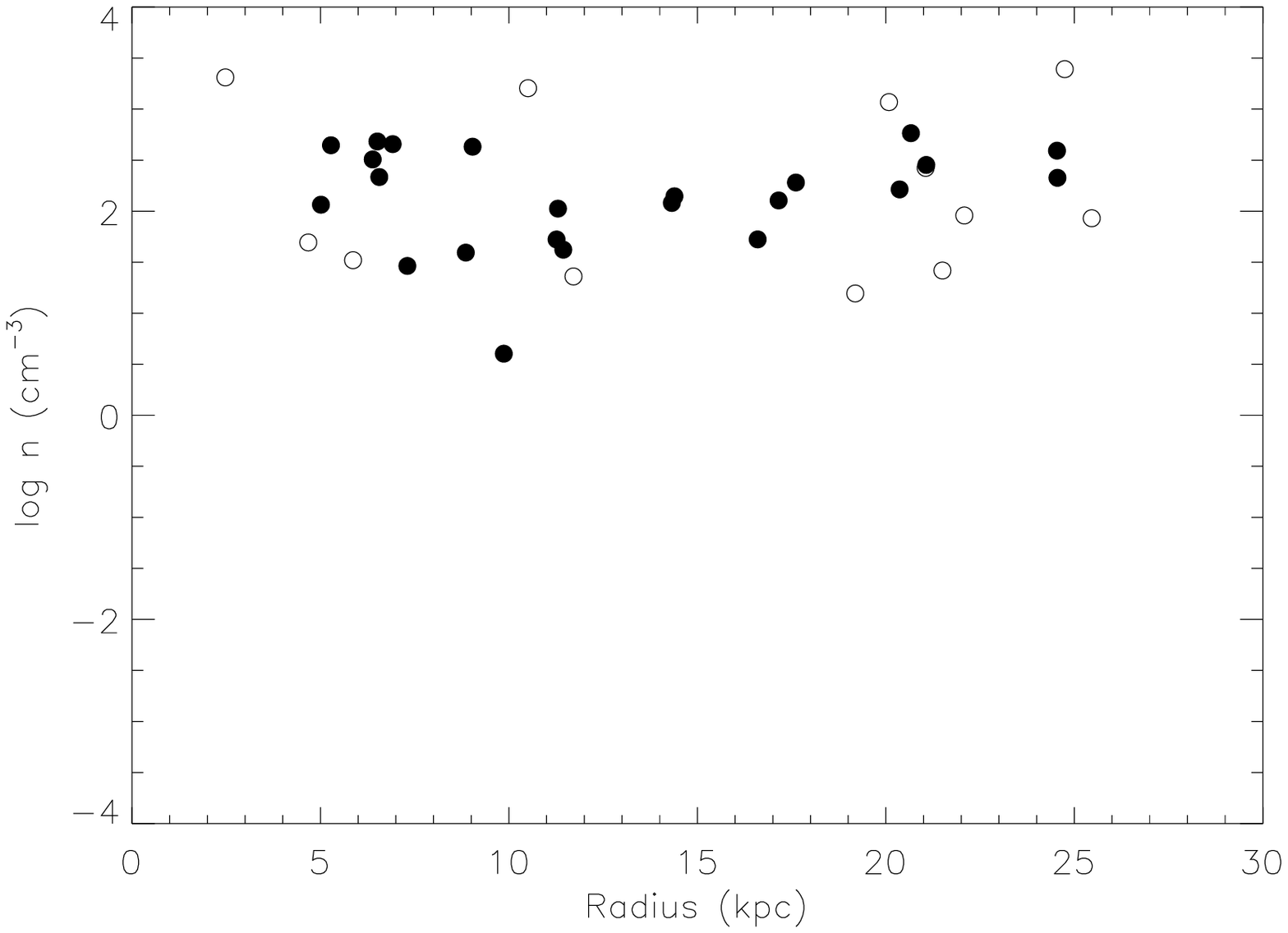}
\figcaption[voltauSC.eps]{Gas Volume Density with varying $A_V/N_H$,
including an extinction correction. The volume densities have been
derived using the Edmunds \& Pagel type metallicity gradient and the Savage
\& Mathis reddening curve. (This is our preferred set of corrections 
at the present time.) \label{fig:tau1}}
\end{figure}

\clearpage
\begin{figure}
\epsscale{0.9}
\plotone{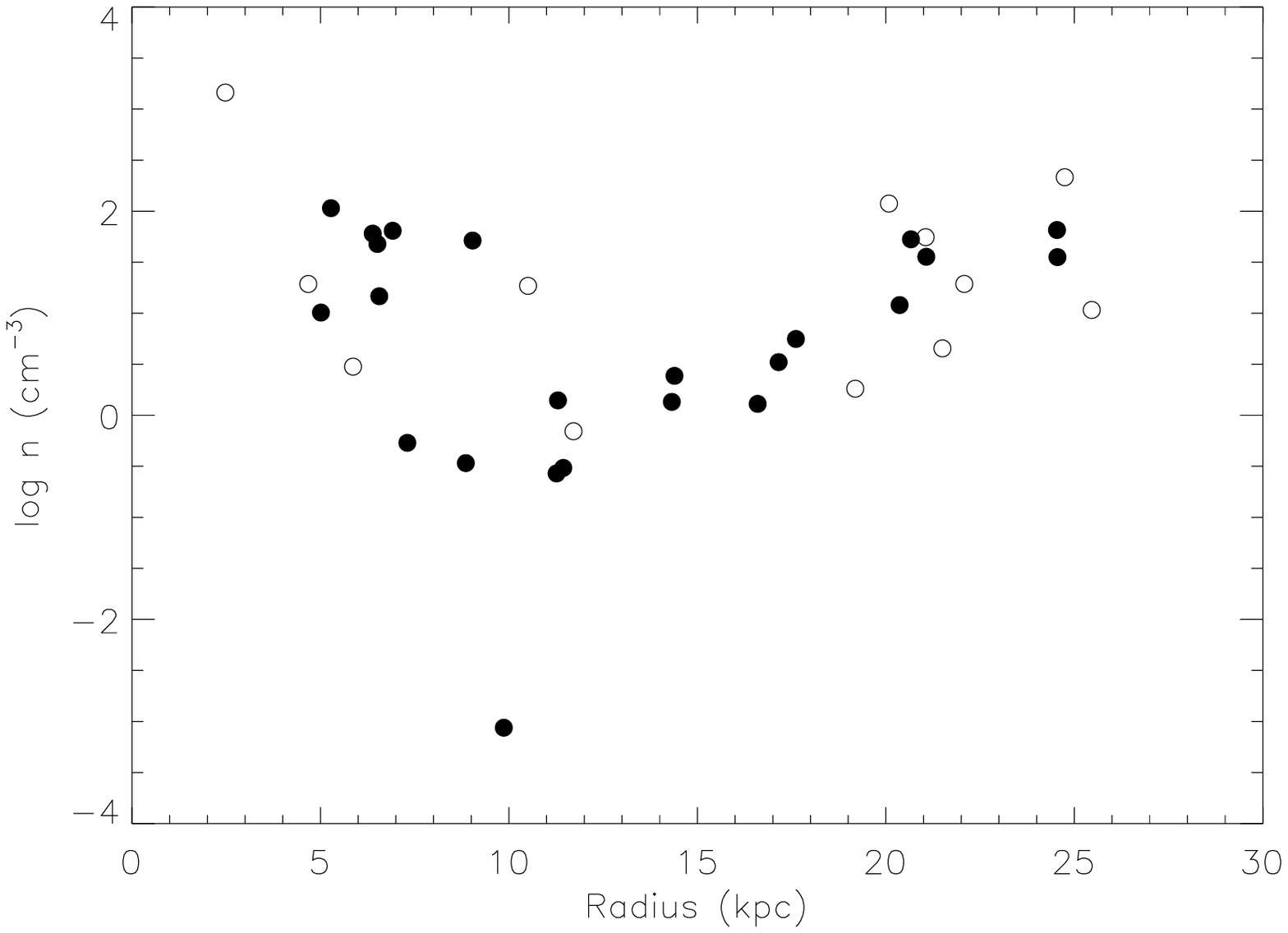}
\figcaption[volZdgtau.eps]{As in Figure \protect \ref{fig:tau1}, but 
with alternative metallicity correction, and including an extinction 
correction. The volume
densities have been derived using the average of the Dopita \& Evans
and MRS calibrations and the Savage \& Mathis reddening
curve. \label{fig:tau2}}
\end{figure}


\clearpage

\begin{deluxetable}{ccc}
\tablecaption{A Sample of H$\alpha$ sources lacking FUV counterparts \label{tab:nofuv}}
\tablewidth{0pt}
\tablehead{
\colhead{Hodge Number} & \colhead{RA (1950)} & \colhead{Dec (1950)}}
\startdata
888 & $14^h 1^m 38.5^s$ & $+54^\circ 38' 34.4''$ \\ 
891 & 14 1 38.3 & $+54$ 35 46.5 \\
939 & 14 1 40.5 & $+54$ 35 10.8 \\
979 & 14 1 42.4 & $+54$ 36 4.8 \\
1117 & 14 1 56.9 & $+54$ 33 44.2 \\ 
1122 & 14 1 58.4 & $+54$ 34 0.1 \\
1126 & 14 1 58.9 & $+54$ 34 7.0 \\
\enddata
\end{deluxetable}

\begin{deluxetable}{rrrrrrrrrr}
\tabletypesize{\scriptsize}
\tablecaption{Observed and Derived Properties of the 35 Candidate PDRs. \label{tab:data}}
\tablewidth{0pt}
\tablehead{
\colhead{RA}   & \colhead{Dec}   & \colhead{Radius} & \colhead{$F_{FUV}$} &
\colhead{$\rho_{HI}$}  & \colhead{$\chi$} & \colhead{$N_{HI}$} &
\colhead{$\rho_{H\alpha}$}  &
\colhead{$n_{raw}$\tablenotemark{a}}   & \colhead{$n_{corr}$\tablenotemark{b}} \\
\colhead{(1950)} & \colhead{(1950)} & \colhead{(kpc)} & \colhead{(\UITunits)} & 
\colhead{(pc)}   & \colhead{} & \colhead{(atoms cm$^{-2}$)} & \colhead{(pc)} & 
\colhead{(cm$^{-3}$)} & \colhead{(cm$^{-3}$)} 
}
\startdata
 $14^h  1^m 18.77^s$  &    $+54^\circ 35'  26.9''$  &   2.5 & $   3.00 \times 10^{-15}$ &  359. &   0.48 & $  0.23 \times 10^{21}$ & 225. &   76. &   2047. \\
 14  1 13.93  &    $+$54 36  24.9  &   4.7 & $   1.21 \times 10^{-15}$  &  682. &   0.05 & $  0.53 \times 10^{21}$ & 110. &   2.6 &     49. \\
 14  1 42.47  &    $+$54 34  52.9  &   5.0 & $   5.99 \times 10^{-15}$  &  144. &   6.03 & $  1.73 \times 10^{21}$ & 109. &   18. &    116. \\
 14  1 12.11  &    $+$54 33  58.9  &   5.3 & $   1.52 \times 10^{-15}$  &  144. &   1.53 & $  0.89 \times 10^{21}$ & 107. &   28. &    443. \\
 14  1 31.43  &    $+$54 37  52.9  &   5.9 & $   6.32 \times 10^{-15}$  &  430. &   0.71 & $  1.58 \times 10^{21}$ & 106. &   2.8 &     33. \\
 14  1 26.37  &    $+$54 32  16.9  &   6.4 & $  14.27 \times 10^{-15}$  &  502. &   1.17 & $  0.97 \times 10^{21}$ & 168. &   18. &    322. \\
 14  1 08.87  &    $+$54 36  50.9  &   6.5 & $   7.21 \times 10^{-15}$  &  161. &   5.74 & $  1.45 \times 10^{21}$ & 147. &   30. &    482. \\
 14  1 47.07  &    $+$54 34  32.8  &   6.6 & $   9.45 \times 10^{-15}$  &  201. &   4.85 & $  1.73 \times 10^{21}$ & 147. &   14. &    216. \\
 14  1 13.24  &    $+$54 37  48.9  &   6.9 & $   4.58 \times 10^{-15}$  &  201. &   2.35 & $  1.15 \times 10^{21}$ & 167. &   24. &    454. \\
 14  1 10.04  &    $+$54 32  50.9  &   7.3 & $  18.47 \times 10^{-15}$  &  359. &   2.98 & $  2.61 \times 10^{21}$ & 314. &   1.5 &     29. \\
 14  1 39.00  &    $+$54 31  32.9  &   8.9 & $   8.69 \times 10^{-15}$  &  215. &   3.89 & $  3.12 \times 10^{21}$ & 240. &  0.68 &     39. \\
 14  0 58.07  &    $+$54 34  30.7  &   9.0 & $   3.32 \times 10^{-15}$  &  201. &   1.71 & $  1.15 \times 10^{21}$ & 110. &   17. &    429. \\
 14  1 55.57  &    $+$54 33  24.7  &   9.9 & $  31.81 \times 10^{-15}$  &  287. &   8.01 & $  5.86 \times 10^{21}$ & 118. &0.0059 &      4. \\
 14  1 37.89  &    $+$54 39  50.9  &  10.5 & $  15.98 \times 10^{-15}$  &   72. &  64.38 & $  3.00 \times 10^{21}$ & 120. &   15. &   1604. \\
 14  2 02.25  &    $+$54 34  18.6  &  11.3 & $   3.20 \times 10^{-15}$  &  144. &   3.22 & $  2.95 \times 10^{21}$ &  74. &  0.79 &    106. \\
 14  1 56.54  &    $+$54 38  10.7  &  11.3 & $   3.48 \times 10^{-15}$  &  144. &   3.50 & $  3.70 \times 10^{21}$ & 126. &  0.19 &     53. \\
 14  0 50.46  &    $+$54 36  10.6  &  11.4 & $  15.13 \times 10^{-15}$  &  395. &   2.02 & $  3.44 \times 10^{21}$ & 114. &  0.19 &     42. \\
 14  1 55.40  &    $+$54 38  44.7  &  11.7 & $   1.46 \times 10^{-15}$  &  305. &   0.33 & $  2.29 \times 10^{21}$ & 144. &  0.30 &     23. \\
 14  2 11.26  &    $+$54 36  48.4  &  14.3 & $   4.07 \times 10^{-15}$  &  108. &   7.28 & $  3.92 \times 10^{21}$ &  77. &  0.26 &    120. \\
 14  1 14.89  &    $+$54 28  46.9  &  14.4 & $  38.71 \times 10^{-15}$  &  359. &   6.24 & $  3.48 \times 10^{21}$ & 106. &  0.53 &    140. \\
 14  0 33.69  &    $+$54 34  24.2  &  16.6 & $  10.44 \times 10^{-15}$  &  305. &   2.33 & $  4.02 \times 10^{21}$ &  80. & 0.068 &     53. \\
 14  0 34.20  &    $+$54 32  12.2  &  17.6 & $  16.23 \times 10^{-15}$  &  215. &   7.27 & $  4.23 \times 10^{21}$ &  90. &  0.14 &    191. \\
 14  0 34.08  &    $+$54 37  38.2  &  17.2 & $   2.98 \times 10^{-15}$  &  108. &   5.35 & $  4.19 \times 10^{21}$ & 131. &  0.11 &    127. \\
 14  0 25.65  &    $+$54 34  02.0  &  19.2 & $   2.34 \times 10^{-15}$  &  610. &   0.13 & $  2.29 \times 10^{21}$ & 139. &  0.12 &     16. \\
 14  0 24.30  &    $+$54 32  53.9  &  20.1 & $   3.00 \times 10^{-15}$  &   72. &  12.10 & $  2.91 \times 10^{21}$ & 145. &   3.2 &   1172. \\
 14  0 53.12  &    $+$54 43  22.7  &  20.4 & $  10.09 \times 10^{-15}$  &  287. &   2.54 & $  3.79 \times 10^{21}$ & 142. &  0.12 &    164. \\
 14  2 25.61  &    $+$54 39  38.1  &  20.7 & $  27.62 \times 10^{-15}$  &  287. &   6.96 & $  3.40 \times 10^{21}$ & 111. &  0.70 &    581. \\
 14  0 40.94  &    $+$54 28  02.4  &  21.1 & $   2.68 \times 10^{-15}$  &  161. &   2.14 & $  2.69 \times 10^{21}$ & 100. &  0.88 &    284. \\
 14  0 29.87  &    $+$54 40  34.1  &  21.1 & $  10.78 \times 10^{-15}$  &  610. &   0.60 & $  1.11 \times 10^{21}$ & 640. &   6.7 &    266. \\
 14  1 56.62  &    $+$54 44  14.7  &  21.5 & $   1.12 \times 10^{-15}$  &  466. &   0.11 & $  1.86 \times 10^{21}$ & 139. &  0.24 &     26. \\
 14  0 22.74  &    $+$54 30  49.9  &  22.1 & $   1.51 \times 10^{-15}$  &  395. &   0.20 & $  1.24 \times 10^{21}$ & 142. &   1.7 &     91. \\
 14  2 30.28  &    $+$54 41  57.9  &  24.5 & $   2.30 \times 10^{-15}$  &  161. &   1.83 & $  3.00 \times 10^{21}$ & 196. &  0.41 &    392. \\
 14  1 50.63  &    $+$54 46  06.8  &  24.6 & $   3.93 \times 10^{-15}$  &  287. &   0.99 & $  3.00 \times 10^{21}$ & 188. &  0.22 &    212. \\
 14  2 43.54  &    $+$54 38  05.4  &  24.7 & $  98.54 \times 10^{-15}$  &  251. &  32.41 & $  5.77 \times 10^{21}$ &  17. & 0.028 &   2466. \\
 14  1 05.56  &    $+$54 46  40.8  &  25.5 & $   2.04 \times 10^{-15}$  &  251. &   0.67 & $  4.71 \times 10^{21}$ & 268. &0.0049 &     85. \\
 \enddata


\tablenotetext{a}{Volume densities derived using the (unrealistic) case of $A_V/N_H = \mbox{constant}$.}
\tablenotetext{b}{Volume densities derived using the Edmunds \& Pagel metallicity gradient and Savage \& 
Mathis extinction law.}


\end{deluxetable}

\end{document}